\documentclass[useAMS,usenatbib]{mnras}
\usepackage{amssymb}
\usepackage{graphicx}
\usepackage{xspace}
\usepackage{amssymb,latexsym,graphicx,natbib,eufrak,times,amsmath}

\usepackage{ifthen}
\def\draftversion{1} 

\makeatletter
\newcommand\mytoc{%
    \@starttoc{toc}%
}
\makeatother

\ifthenelse{\equal{\draftversion}{1}}{
	\usepackage{xcolor}
	\newcommand{\tmp}{}
	\newenvironment{envcomm}[1]{\renewcommand{\tmp}{#1}\begin{color}[rgb]{0,0.5,0.0}\begin{center}\hrule\vspace{0.5mm}\tmp's COMMENTS\end{center}}{\begin{center}END OF \tmp's COMMENTS\vspace{0.5mm}\hrule\end{center}\end{color}}
	\newenvironment{draft}{\begin{color}[rgb]{0,0.4,0}\begin{center}\hrule\vspace{0.5mm}DRAFT\end{center}}{\begin{center}END OF DRAFT\vspace{0.5mm}\hrule\end{center}\end{color}}
	\newcommand{\comcomm}[2]{\begin{color}[rgb]{0,0.5,0.0}\ $\bullet$ \textbf{#1:} #2 $\bullet$\ \end{color}}
	\newcommand{\revend}[1]{\par\begin{color}[rgb]{0,0.4,0}\begin{center}\hrule\vspace{0.5mm}END OF #1's REVISIONS\vspace{0.5mm}\hrule\end{center}\end{color}\par}
	\newcommand{\todo}[1]{\begin{color}{red}\ $\bullet$ \textbf{To do: }#1 $\bullet$\ \end{color}}

	\newcommand{\del}[1]{\begin{color}[rgb]{0,0.5,0.0}\ $\bullet$ \textbf{Deleted: }#1 $\bullet$\ \end{color}}
	\newcommand{\sk}[1]{\begin{color}[rgb]{0.6,0,0.6}#1\end{color}}
	\newcommand{\toc}{\par\begin{color}[rgb]{0.6,0,0.6}\begin{center}\hrule\vspace{0.5mm}\begingroup\small\let\cleardoublepage\relax\let\clearpage\relax\mytoc\endgroup\vspace{0.5mm}\hrule\end{center}\end{color}\par}
	}{
	\newsavebox{\trashcan}
	\newenvironment{envcomm}[1]{\begin{lrbox}{\trashcan}\begin{minipage}{\columnwidth}}{\end{minipage}\end{lrbox}}
	
	\newcommand{\comcomm}[2]{}
	\newcommand{\revend}[1]{}
	\newcommand{\todo}[1]{}

	\newcommand{\del}[1]{}
	\newcommand{\sk}[1]{}
	\newcommand{\toc}{}
	}

\long\def\symbolfootnote[#1]#2{\begingroup%
\def\thefootnote{\fnsymbol{footnote}}\footnote[#1]{#2}\endgroup} 

\newcommand{\fig}[2][]{Figure#1~\ref{fig:#2}}
\newcommand{\tab}[2][]{Table#1~\ref{tab:#2}}
\newcommand{\sect}[2][]{Section#1~\ref{sec:#2}}

\renewcommand{\fig}[2][]{Fig#1.~\ref{fig:#2}}

\newcommand{\mh}{\ensuremath{\textrm{\,--\,}}}
\newcommand{\bb}[1]{\ifmmode \mbox{\boldmath $ #1$} \else  \mbox{\boldmath $#1$} \fi}

\newcommand{\U}[1]{\ensuremath{\mathrm{~#1}}}
\newcommand{\e}[1]{\ensuremath{\times 10^{#1}}}
\newcommand{\yr}{\U{yr}}
\newcommand{\Myr}{\U{Myr}}
\newcommand{\Gyr}{\U{Gyr}}
\newcommand{\pc}{\U{pc}}
\newcommand{\kpc}{\U{kpc}}

\newcommand{\msun}{\U{M}_{\odot}}
\newcommand{\Msun}{\msun}
\newcommand{\Msunyr}{\Msun\yr^{-1}}

\newcommand{\cc}{\U{cm^{-3}}}

\newcommand{\kms}{\U{km\ s^{-1}}}

\newcommand{\hi}{H{\sc i}\xspace}
\newcommand{\hii}{H{\sc ii}\xspace}
\newcommand{\ha}{H\ensuremath{\alpha}\xspace}

\newcommand{\qt}{\ensuremath{Q_\mathrm{T}}}

\newcommand{\ramses}{{\small RAMSES}\xspace}

\newcommand{\hop}{{\small HOP}\xspace}

\title[Star formation in the Cartwheel]{Morphology and enhanced star formation in a Cartwheel-like ring galaxy}

\author[Renaud et al.] {F.~Renaud$^{1}$\thanks{f.renaud@surrey.ac.uk}, E.~Athanassoula$^{2}$, P.~Amram$^{2}$, A.~Bosma$^{2}$, F.~Bournaud$^{3}$, P.-A.~Duc$^{3,4}$,\newauthor B.~Epinat$^{2}$, J.~Fensch$^{3}$, K.~Kraljic$^{2}$, V.~Perret$^{5}$, C.~Struck$^{6}$\\
$^1$ Department of Physics, University of Surrey, Guildford, GU2 7XH, UK\\
$^2$ Aix Marseille Univ., CNRS, LAM, Laboratoire d'Astrophysique de Marseille, 13388 Marseille, France\\
$^3$ Laboratoire AIM Paris-Saclay, CEA/IRFU/SAp, Universit\'e Paris Diderot, F-91191 Gif-sur-Yvette Cedex, France\\
$^4$ Observatoire Astronomique de Strasbourg, Universit\'e de Strasbourg, CNRS UMR 7550, 11 rue de l'Universit\'é, F-67000 Strasbourg, France\\
$^{5}$ Institute for Theoretical Physics, University of Z\"urich, CH-8057 Z\"urich, Switzerland\\
$^{6}$ Department of Physics and Astronomy, Iowa State University, Ames, IA 50011, USA
}

\date{Accepted 2017 September 7. Received 2017 September 1; in original form 2017 June 26}

\begin{document}
\maketitle


\begin{abstract}
We use hydrodynamical simulations of a Cartwheel-like ring galaxy, modelled as a nearly head-on collision of a small companion with a larger disc galaxy, to probe the evolution of the gaseous structures and flows, and to explore the physical conditions setting the star formation activity. Star formation is first quenched by tides as the companion approaches, before being enhanced shortly after the collision. The ring ploughs the disc material as it radially extends, and almost simultaneously depletes its stellar and gaseous reservoir into the central region, through the spokes, and finally dissolve 200 Myr after the collision. Most of star formation first occurs in the ring before this activity is transferred to the spokes and then the nucleus. We thus propose that the location of star formation traces the dynamical stage of ring galaxies, and could help constrain their star formation histories. The ring hosts tidal compression associated with strong turbulence. This compression yields an azimuthal asymmetry, with maxima reached in the side furthest away from the nucleus, which matches the star formation activity distribution in our models and in observed ring systems. The interaction triggers the formation of star clusters significantly more massive than before the collision, but less numerous than in more classical galaxy interactions. The peculiar geometry of Cartwheel-like objects thus yields a star (cluster) formation activity comparable to other interacting objects, but with notable second order differences in the nature of turbulence, the enhancement of the star formation rate, and the number of massive clusters formed.
\end{abstract}


\section{Introduction}

In the spectacular zoo of interacting galaxies, collisional rings are one of the most peculiar species. \citet{Lynds1976}, \citet{Theys1977} and \citet{Toomre1978} showed that rings result from the head-on high-speed collision of a companion galaxy on a target disc galaxy. The gravitational acceleration induced by the companion first drags the matter of the target toward the (almost) central impact point. As the intruder flies away, the disc material rebounds, and forms an annular over-density that travels radially. Therefore, an empty region appears behind the ring but in some cases, a nucleus near the impact point can survive. Secondary, inner rings, in the outskirts of nuclear region can then form from second rebounds \citep[see][for reviews]{Athanassoula1985, Appleton1996}. Gas flows onto the nuclear region and inner ring and fuels star formation there \citep{Struck1987, Struck1996}. Collisions with a more off-centred impact lead to less symmetric configurations and could explain why some ring systems do not show a nuclear component, which could also result from the destruction of a loosely bound nucleus during the interaction \citep{Madore2009}. The densest structures like the main ring and the inner ring often host a strong star formation activity. Such activity is rapidly evolving in time and space, as suggested by colour gradients across the structures \citep{Marcum1992, Appleton1997} and as modelled by e.g. \citet{Struck1987}.

Their peculiar morphology make ring galaxies more easily identifiable in observations than other interacting systems. Therefore, rings are often used as probes to reconstruct the galaxy interaction and merger rates and their dependence with redshift to better constraint the assembly history of galaxies \citep{Lavery2004, DOnghia2008}. However, these galaxies are fast evolving systems, as the ring dissolves or fragments within a few $\sim 100 \Myr$ \citep{Theys1977}, and therefore their detectability is a strong function of time. Dynamical models are thus necessary to pin down the conditions required to form a ring and the possible other structures (inner ring, nucleus) and to time the evolution of such systems as well as that of the observable components (neutral gas, excited gas, dust, star forming regions etc.).

Numerous analytical and numerical models have addressed the questions of the formation of ring galaxies, mostly focussing on the large-scale dynamics of the galaxies, either to constrain the formation scenario of a specific system \citep[e.g.][]{Struck1987, Gerber1992, Struck1993, Hernquist1993, Bosma2000, Horellou2001, Vorobyov2003, Bournaud2007, Ghosh2008, Mapelli2008, Michel2010, Smith2012}, or to explore the parameter space leading to the formation of ring-like systems \citep[e.g.][]{Huang1988, Appleton1990, Struck1990, Gerber1994, Gerber1996, Athanassoula1997, Romano2008, DOnghia2008, Mapelli2012, Fiacconi2012}. Such approach calls for a large number of simulations either in surveys or with a trial-and-error method, at a high numerical cost. Therefore, resolution and/or the range of physical processes captured are most often sacrificed. However, modelling the physics of star formation and the associated feedback, both key in the evolution of galaxies and their structures, requires to capture the organisation of the dense phase of the interstellar medium (ISM), which goes through resolving the turbulent cascade from its injection scale(s) down to, at least, its supersonic regime. \citet{Renaud2014b} showed that, in the case of galaxy interaction driven turbulence, this regime and thus the convergence of the star formation rate (SFR) are reached at the scale of a few parsecs ($\sim 1\mh 10 \pc$).

Here, we present models of a collisional ring system at this resolution and explore the dynamics and hydrodynamics ruling the organisation of the ISM in molecular clouds ($\sim 10\mh 100 \pc$) and the associated star formation activity, and their evolution in time and space. To do so, we focus on the particular case of Cartwheel-like galaxies.

Discovered by \citet{Zwicky1941}, the Cartwheel galaxy is part of the compact group of 4 galaxies SCG~0035-3357 (see \citealt{Iovino2002} and e.g. the figure 1 of \citealt{Bosma2000} for identification). The companion responsible for the formation of the ring has not yet been unambiguously identified, but the presence of a long \hi plume between the ring galaxy and the companion G3 \citep{Higdon1996} is a robust hint of tidal interaction. Extended distributions of neutral gas around galaxies are easily tidally disturbed during interactions to form bridges between galaxies and tails, that constitute long-lived signatures of past encounters (see \citealt{Toomre1972}, and \citealt{Duc2013} for a review). The long \hi plume in the Cartwheel system would thus be the bridge originating from G3 and extending up to the ring galaxy itself. Ram pressure stripping could also play a role in the formation of this plume, as a fraction of the G3 gas would be stripped as the galaxy flies in the hot halo of the Cartwheel.

The Cartwheel galaxy encompasses a wide range of physical conditions (ring, spokes, inner ring, nucleus), each harbouring a different star formation regime. For instance, the ring structure contains most of the \hi gas of the galaxy \citep{Higdon1996}, hosts star forming regions \citep{Marcum1992, Fosbury1977, Higdon1995, Amram1998, Charmandaris1999, Wolter1999, Higdon2015} and ultra-luminous X-ray sources \citep{Gao2003, Wolter2004}. These points indicate an enhanced star formation activity in the form of massive stars and star clusters in this region. However, \citet{Amram1998} reported little \ha emission in the spokes and the nucleus, indicating a lower level of star formation there. Furthermore, the nucleus contains hot dust (traced by mid-infrared, \citealt{Charmandaris1999}) and cold, diffuse gas \citep{Horellou1998}, consistent with a weak star formation activity. Understanding why most star formation occurs in the ring despite large quantities of gas found in the spokes and the nuclear region requires detailed hydrodynamics models capturing the physical processes coupling the galactic-scale dynamics with the star formation activity.

We use the simulations presented here (i) to better understand the (hydro-)dynamics of Cartwheel-like systems and the evolution of their main components, and (ii) as a laboratory to explore the physics of enhanced star formation and the formation of massive star clusters in a (relatively) simple geometry. The goal of this work is not to improve the match between models and observations, but rather to study the formation and evolution of Cartwheel-like objects, i.e. disc galaxies with two rings linked by what is commonly referred to as spokes. More generally, we treat these simulations as numerical experiments in which we probe a series of physical processes over a range of physical conditions in the multi-scale and multi-physics problem of triggered star formation.

\section{Method}
\label{sec:method}

We use the adaptive mesh refinement (AMR) code \ramses \citep{Teyssier2002} and the same methods for heating, cooling, star formation and feedback as in \citet{Renaud2017}, based on \citet{Agertz2015, Agertz2016}, but with the parameters given below. Briefly, star formation occurs in gas denser than $500 \cc$ (while densities are typically probed up to $\sim 10^5\cc$ in our simulations, see below), with an efficiency per free-fall time of $1\%$, which leads for our models to a global SFR of the order of $\sim 1 \Msunyr$ before the collision. Each stellar particle represents a single age population following a \citet{Chabrier2003} initial mass function. Feedback from the massive young stars injects energy, momentum and mass in the interstellar medium, through winds and supernovae (types II and Ia), either in the form of energy or momentum depending on the resolution of the local cooling radius, following \citet{Kim2015}. The cell size reaches $6 \pc$ in the most refined regionss of the $400 \kpc$ wide cubic simulation volume.

\begin{table}
\caption{Initial setup of the main galaxy}
\label{tab:ic}
\begin{tabular}{lc}
\hline
\multicolumn{2}{l}{Gas disc (exponential)}\\
mass [$\times 10^9 \msun$]  & 16.0 \\
radial scale length [kpc] & 6.0  \\
truncation radius [kpc]     & 30.0 \\
vertical scale length [kpc] & 0.2 \\
truncation height [kpc]     & 2.0  \\
\hline
\multicolumn{2}{l}{Stellar disc (exponential)}\\
mass [$\times 10^9 \msun$]          & 48.0 \\
radial scale length [kpc]         & 3.0  \\
truncation radius [kpc]             & 20.0 \\
vertical scale length [kpc]         & 0.1 \\
truncation height [kpc]             & 1.0 \\
\hline
\multicolumn{2}{l}{Stellar bulge (Plummer)}\\
mass [$\times 10^9 \msun$]          & 12.0 \\
characteristic radius [kpc]         & 1.0 \\
truncation radius [kpc]             & 10.0 \\
\hline
\multicolumn{2}{l}{Dark matter halo (Plummer)}\\
mass [$\times 10^9 \msun$]          & 244.0 \\
characteristic radius [kpc]         & 6.0  \\
truncation radius [kpc]             & 60.0  \\
\hline
Total mass [$\times 10^9 \msun$] & 320.0 \\
\hline
\end{tabular}
\end{table}

\begin{table}
\caption{Initial orbital parameters}
\label{tab:ic_orbit}
\begin{tabular}{lcc}
\hline
& Main galaxy & Companion\\
position [kpc] & 0,0,0 & -0.71, 13.2, 52.6\\
velocity [km/s] & 0,0,0 & 13.6, -94.8, -379.4\\
spin axis & 0,0,1 & - \\
\hline
\end{tabular}
\end{table}

We model the target galaxy as a disc+bulge+halo galaxy and add one impacting high-speed companion which induces the tidal perturbations on the main galaxy and thus forms the ring. For the sake of simplicity, we do not consider the other galaxy members of the SCG~0035-3357 compact group. The internal structure of the companion is not analysed here and thus we render it with only a live dark matter halo, for simplicity. The main galaxy and its companion are set up in isolation, neglecting the cosmological context, with intrinsic parameters largely based on \citet{Horellou2001}, designed to reproduce the observed morphology of the system, and given in \tab{ic}. The stellar disc, bulge and halo of the main galaxy are initially rendered with 300,000, 75,000 and 135,000 particles respectively to model the (live) gravitational potential of the galaxy. The gas and the stars formed during the simulation are treated at the local resolution of the AMR grid, i.e. at a much higher mass and space resolution. The companion is rendered with 80,000 particles following a \citet{Plummer1911} distribution with a scale radius of $3 \kpc$, a total mass of $160\e{9} \Msun$ and a mass ratio of 1:2 between the two galaxies. Such mass ratio was initially proposed by \citet{Horellou2001} to match the observed morphology of the system. It is however significantly higher than the ratio derived observationally between the Cartwheel and any other member of the compact group \citep{Higdon1996, Bosma2000}. One can argue that a significant amount of dynamical mass is lost during the encounter, especially for the low-mass, fragile companion, which introduces large differences between the present-day estimated mass and the pre-collision value. Our models confirm that a significant fraction of the companion's mass is spread in the tidal debris after the collision. An estimate of its mass loss would however require to model its baryonic content.

We run three simulations, keeping all intrinsic and orbital parameters strictly identical but varying only the velocity structure of the disc stars in the target to model a galaxy with an initial Toomre stability parameter \citep{Toomre1964} of $\qt = 0.5$, 1 and 2 for the stellar component. The actual value evolves during the simulations as the disc relaxes and forms new stars but the relative differences between the models are approximately conserved. Doing so allows the main disc to form different structures before the collision, which regulates the formation of post-collision features. Different stabilities also influence the star formation activity before and after the collision, as discussed in the next sections.

The orbit is similar to that of \citet{Horellou2001}, but with a larger initial separation between the galaxies (\tab{ic_orbit}), allowing for a longer relaxation of the models and a smoothing out of any imperfection of our initial conditions. In the rest of the paper, $t=0$ corresponds to the instant of the collision (i.e. when the companion's centre of mass is the closest to the mid-plane of the target), which takes place $104 \Myr$ after the beginning of the simulation, $600 \pc$ away from the centre of mass of the target's disc. This slight offset allows to build an off-centred nucleus as well as asymmetric structures, as observed in the Cartwheel galaxy. Along this orbit, the companion crosses very dense regions only for a very short period and thus, the dynamical friction it experiences is too weak to make it lose enough orbital energy to fall back and merge with the target galaxy. The final state of our simulations indicates that the two galaxies are not bound to each other, and thus that the encounter is a unique event.

\section{Results}
\label{sec:results}

\subsection{Overall morphology and evolution}
\label{sec:morpho}

\begin{figure*}
\includegraphics{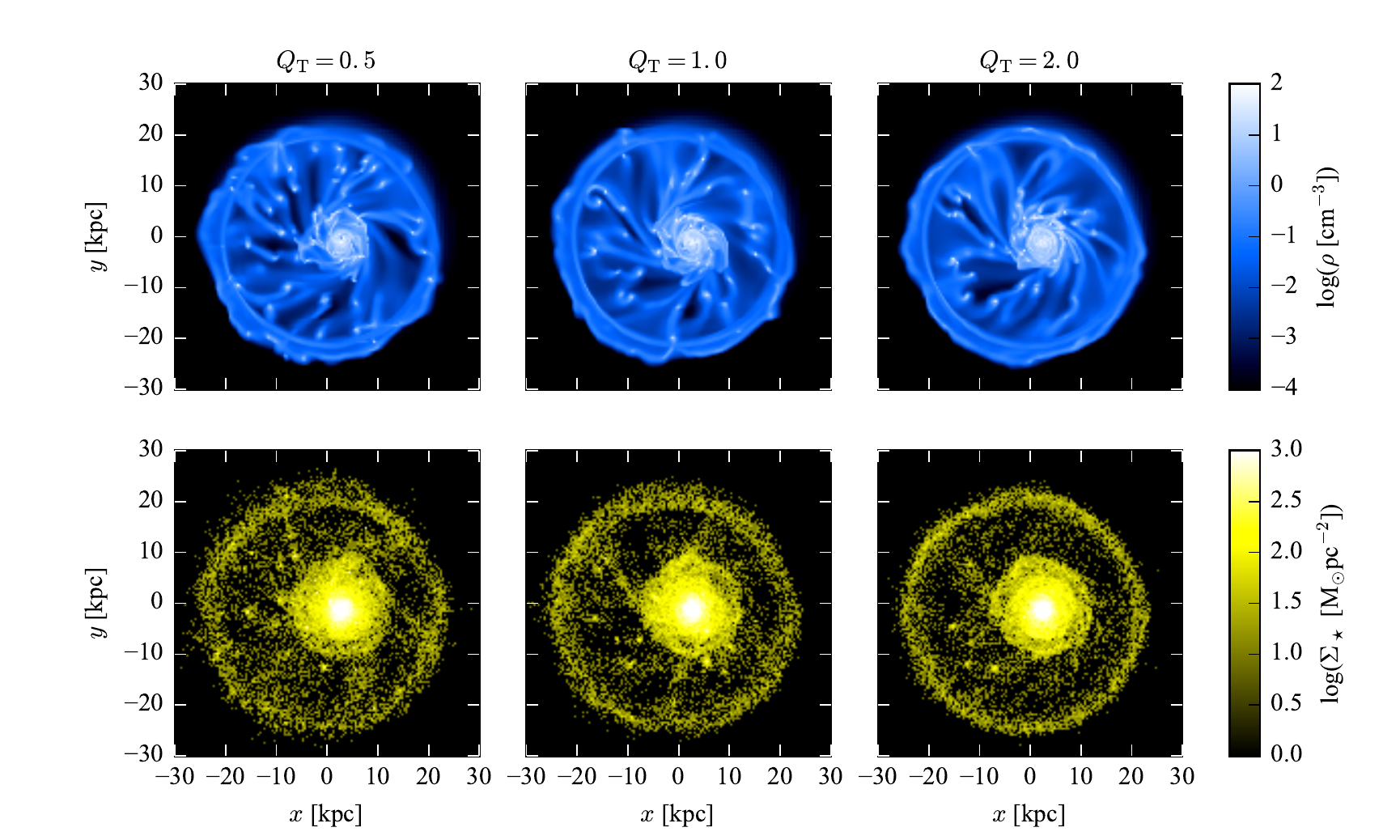}
\caption{Map of the gas (top) and stellar (bottom) density of the main galaxy, $150 \Myr$ after the collision.}
\label{fig:morpho}
\end{figure*}

The mophological and kinematical analysis of the large-scale ($\gtrsim 200 \pc$) structures presented in this section is made based on our $\qt = 1$ model. Similar results are found using the other models as only the small-scale structures and the physics of star formation is affected by the initial stability of the progenitor.\footnote{Several of the structures and their evolution discussed in this paper are illustrated in movies available here: \url{http://www.astro.lu.se/~florent/movies.php}.} This is in agreement with the results of \citet{Athanassoula1997}, who found only a relatively small difference in the density in the spokes for different $\qt$ values. 

Following \citet{Athanassoula1997} and \citet{Horellou2001}, our simulations reproduce the overall morphology of the Cartwheel galaxy (\fig{morpho}). Only a few Myr after the collision, the ring expands and sweeps the disc material of the progenitor. During the first $\approx 200 \Myr$ after the collision, the ring expands radially at the roughly constant speed of $\approx 120 \kms$, making a shock front. This expansion is almost planar (but note the presence of a three-dimensional structure, see \sect{cylinder}). Conversely, the matter close to the impact point gathers in a roughly spherical stellar component (inherited from the bulge) and in a rather thin, planar feature, with a high gas fraction. A secondary, inner ring made of gas and stars is found in the outskirts of this nuclear region \citep[see][]{Athanassoula1997}, and mostly visible in our $\qt=1$ and $\qt=2$ cases. The density contrast between the inner ring and the material it encloses is much lower than that for the main ring, which makes this inner ring more difficult to robustly identify. For this reason, we refer to the inner ring and all it encloses as the nuclear region, without telling them apart.

The large-scale dynamics ($\gtrsim 200 \pc$) and the morphology of main features are comparable in our three cases: the orbit of the companion, the nuclear region, the ring and the area in between that we refer to as the spoke region, for simplicity. The relatively simple geometry of these structures allows us to identify them through a handful of elements, measured by visual inspection of radial density profiles and computation of the centre of mass: the position of the nucleus with respect to the ring's centre, its size and the outer and inner radii of the ring. We use these parameters to tell apart the three main regions in the rest of the paper. During the $\approx 200 \Myr$ following the collision, their time evolution can then be approximately parametrized with linear functions of time.

\begin{figure}
\includegraphics{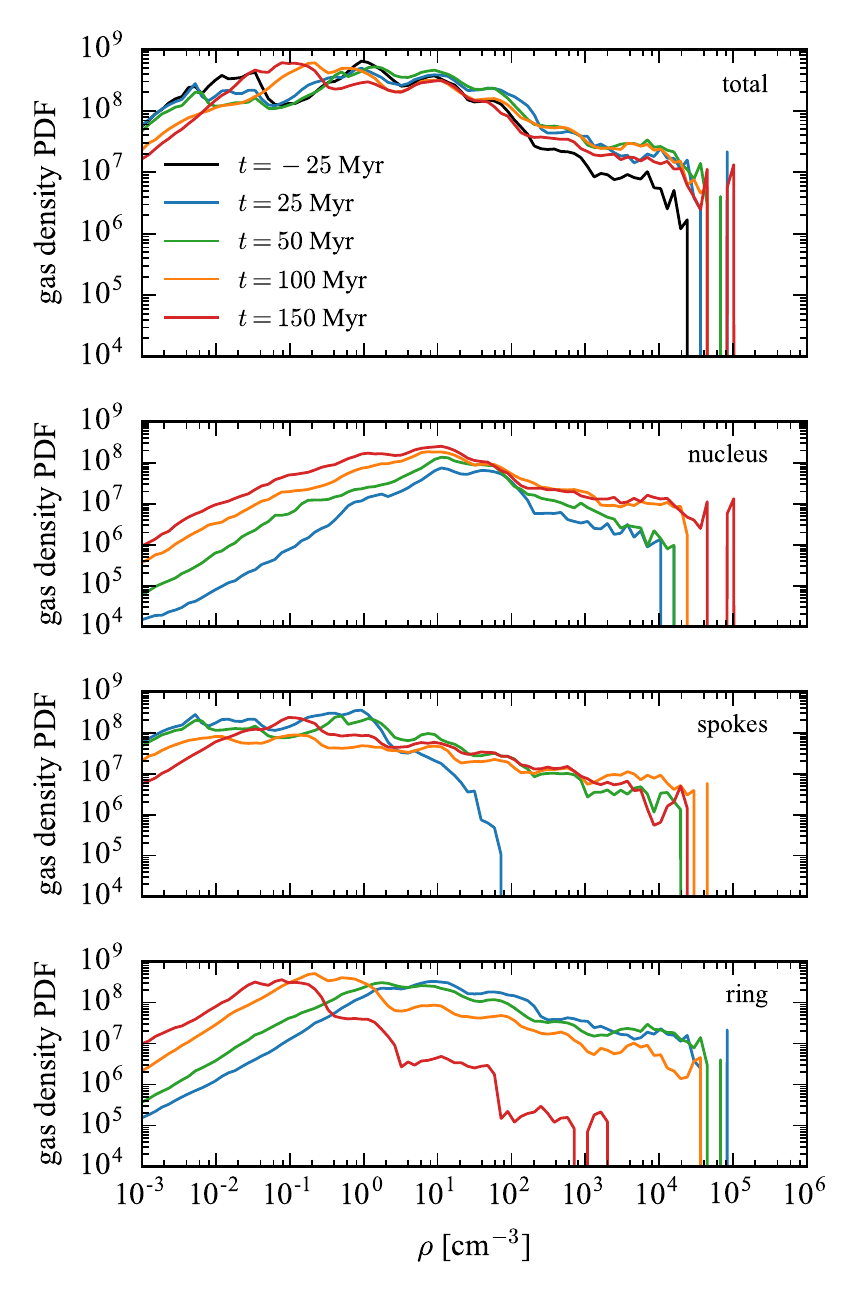}
\caption{Evolution of the mass-weighted gas density PDF in our $\qt =1$ galaxy (top), and in the main three regions of the galaxy. (The black curve representing $t=-25\Myr$ is not shown for the three regions as they are not defined before the collision.)}
\label{fig:pdf}
\end{figure}

\begin{figure}
\includegraphics{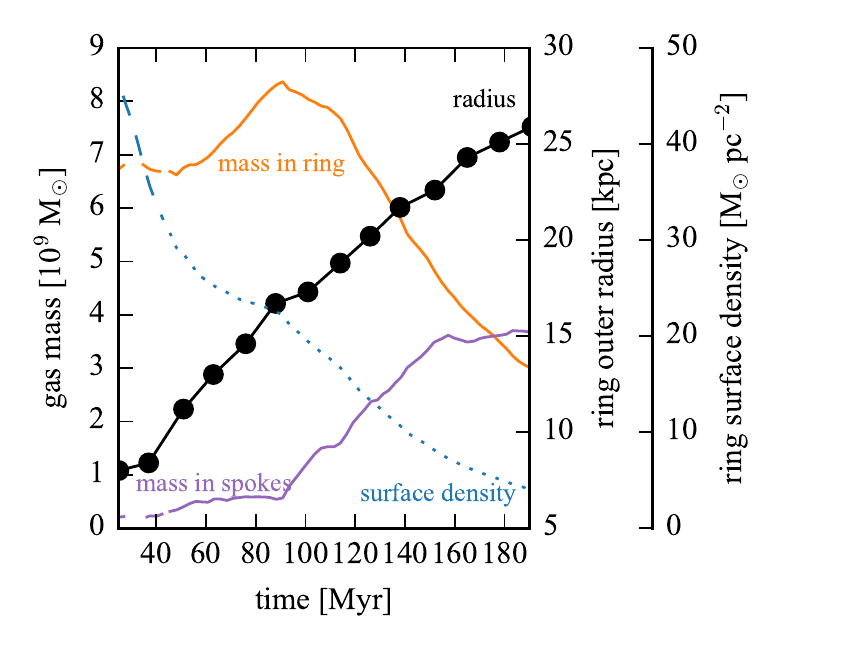}
\caption{Evolution of the gas mass in the main ring (orange) and in the spoke region (purple), of the ring outer radius estimated by eye on planar density profiles (black), and of the ring surface density (dotted blue) in our $\qt =1$ galaxy. Identifying the ring structure is challenging before and after the time-lapse shown here. The dashed part of the curves corresponds to the early epoch of the detection of the ring when it is complicated to define its inner radius, which thus introduces uncertainties in its mass and surface density.}
\label{fig:ringmass}
\end{figure}

\fig{pdf} shows the mass-weighted gas density probability distribution function (PDF) at selected epochs over the entire galaxy, and by telling apart the main regions of the system. The galaxy experiences a significant increase of its maximum density as well as its amount of dense gas shortly after the interaction. Thereafter, the evolution of the PDF is globally milder, but corresponds instead to strong variations and gas flows between the regions of the galaxy.

A few Myr after the collision, a density enhancement expands radially and ploughs through the disc material, which forms the main ring. As the ring propagates, its gathers more and more material (gas and stars) and its mass increases, as illustrated on \fig{ringmass}. While the PDF of the nuclear region remains mostly unaffected by the collision ($25 \mh 50 \Myr$), the regions between the nucleus and the ring (that we call ``spokes'') become relatively empty, in a similar manner as the interior of a blast wave.

As the ring expands, it starts losing material that gradually falls back onto the nucleus by streaming along the spokes. The mass loss overcomes the accumulation of matter when the ring reaches the low density outskirts of the progenitor disc and thus gathers very little amounts of matter (\fig{ringmass}). This happens at $t \approx 90\Myr$, when the ring’s outer radius is $\approx 17 \kpc$, i.e. about 3 times the initial scale-length of the gas disc. At this point, the PDF of the ring yields its first major evolution, with the depletion of the dense tail. As soon as some material, including dense clumps, leaves the ring through the spokes (i.e. as early as $t \approx 25\mh 50\Myr$), the PDF of the spokes develops a tail at high density.

The (hydro)dynamics of the spoke region lower the maximum density (see \sect{shear}) and a significant amount of diffuse gas is found in between the individual spokes. This situation does not significantly evolve until the ring is completely depleted. The material in the spoke region falls onto the nucleus which thus accretes gas spanning a wide range of densities rather than, for instance, only diffuse gas. The shape of its PDF varies only mildly. However, the nuclear inflow generates an excess of very dense gas, mainly found in the innermost region ($\lesssim 1 \kpc$), and along the secondary, inner ring.

As the mass of the main ring continues to decrease due to the lack of further accretion and the depletion through the spokes, the ring completely dissolves $\approx 200 \Myr$ after the collision.

The density distribution of the system thus follows a rather simple global mechanism of compaction in the ring, depletion through the spokes and accretion in the nuclear region (with additional compaction in the inner ring and the very centre).

\subsection{Vertical structure and cylindrical shape}
\label{sec:cylinder}
 
In this section, we ignore the slight radial offset of the companion's trajectory with respect to the spin axis of the main disc and consider, for the sake of simplicity, that the impact point is at the centre of the disc. The actual difference has only second-order effects on the reasoning and results presented here. In our setup, the companion travels from $z>0$ to $z<0$.

\begin{figure}
\includegraphics{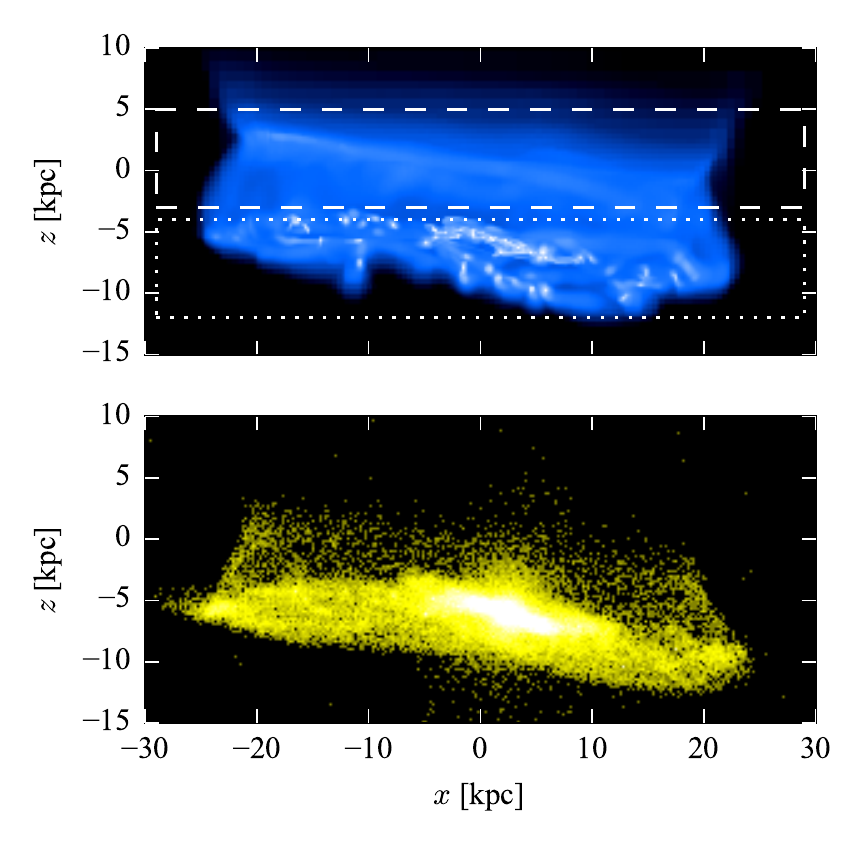}
\caption{Gas (top) and star (bottom) densities at $t=150 \Myr$ in the edge-on view of the main galaxy. Colours scale as in \fig{morpho}. Rectangles in dashed and dotted lines indicate the position of the slices shown in \fig{cylinderslice}.}
\label{fig:cylindery}
\end{figure}

\begin{figure}
\includegraphics{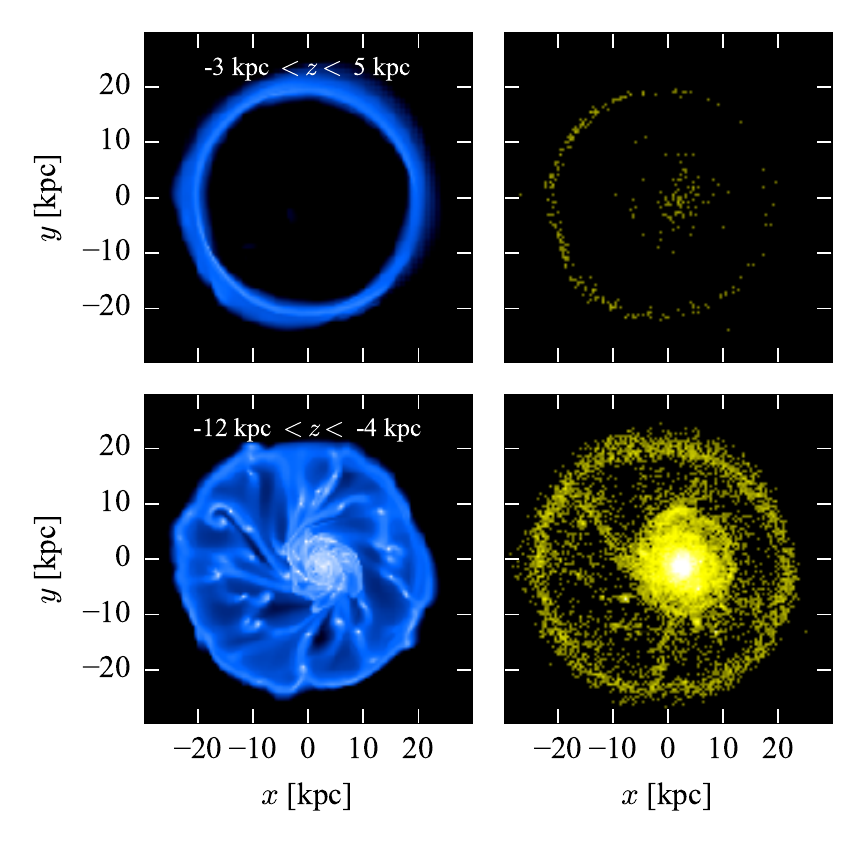}
\caption{Gas (left) and star (right) densities at $t=150 \Myr$ in slices $8 \kpc$ deep, along the $z$-axis and centred on $z=1\kpc$ (top) and $z=-8 \kpc$ (bottom), i.e. on the top and bottom parts of the cylindrical annular structure (see rectangles in \fig{cylindery}). Colours scale as in \fig{morpho}.}
\label{fig:cylinderslice}
\end{figure}

\begin{figure*}
\includegraphics{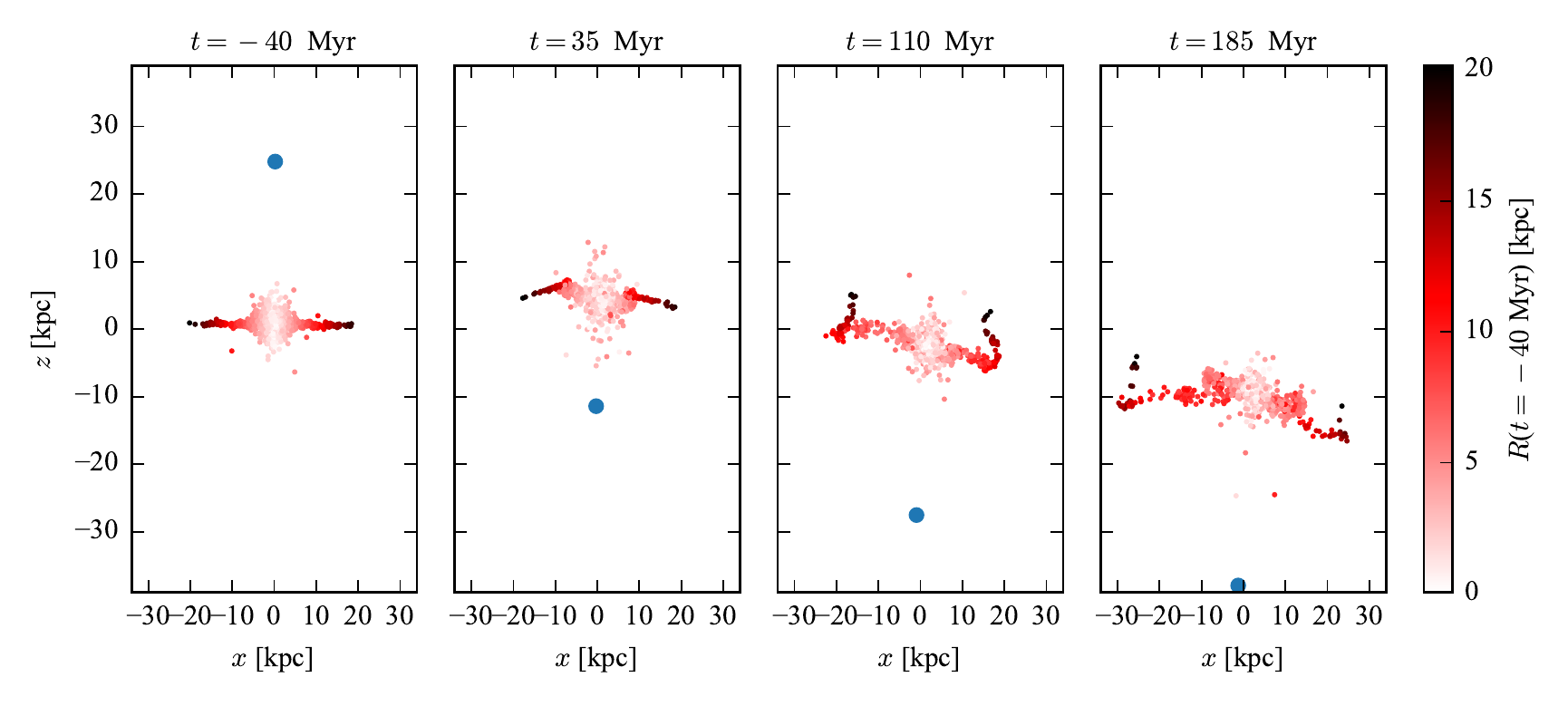}
\caption{Slice view of the position of the stars (in a slice $4 \kpc$ deep, along the $y$-axis). The colour codes the radius (in the $x-y$ plane) of the stars at $t=-40 \Myr$, i.e. before the collision. The vertical structure gathers stars initially at large galactocentric radius. Only a randomly selected subset of the stars is shown, for the sake of clarity. The blue dot marks the position of the centre of mass of the companion galaxy.}
\label{fig:cylinder_sequence}
\end{figure*}

We discuss the formation and evolution of a vertical, cylindrical-like, annular structure induced by the tidal perturbation, visible on the edge-on views of \fig{cylindery} and the slice views in \fig{cylinderslice}. This structure corresponds to the thickening of the target galaxy noted by \citet{Bosma2000}.

All along the interaction, the companion exerts a tidal influence on the main galaxy, which becomes mostly visible in the dynamically cold component, i.e. the stellar and gaseous discs. The amplitude of this effect depends on the strength of the gravitational pull from the companion, which (at a given time) depends on the galactocentric radius. This leads to a radial dependence of the vertical component of the velocity of stars and gas, already before the collision: the outermost regions are less affected by the companion and thus lag behind the inner regions. This introduces a vertical structure in the ``disc''. \fig{cylinder_sequence} shows the evolution of this structure by tagging the particles according to their pre-collision galactocentric radius.

After the collision, the main galaxy is attracted in the opposite direction by the receding companion. Once again, the outer material is less affected by this effect and lags behind the mean plane of the disc. In addition, the ring forms and expands towards the outskirts, sweeping most of the disc material (except the innermost region, $R \lesssim 5 \kpc$, which remains in the nuclear region). By the time the ring reaches the outermost zones, the material there lies above the plane of the disc remnant, and is not affected by the ring (which further slows down the mass growth of the ring, discussed in the previous section).

While expanding, the ring gathers material originating from a range of galactic radii, thus with different vertical velocities inherited from past tidal accelerations. Material from the inner region ($5 \kpc \lesssim R \lesssim 10 \kpc$) retains its high (positive) vertical velocity, while that initially further ($10 \kpc \lesssim R \lesssim 15 \kpc$) moves vertically slower. Therefore, the ring thickens vertically.

As a result of the two effects, the outer galaxy features the shape of a cylindrical annulus (\fig[s]{cylindery} and \ref{fig:cylinder_sequence}). As shown in \fig{cylinderslice}, the bottom surface of the cylinder corresponds to the mean plane of the galaxy and gathers most of the gaseous and stellar material, the ring, the spokes, the nuclear region and all clumps and clusters. At the top of the cylinder the matter is distributed in a thin ring. Inside the ring, the inner region only hosts a tiny fraction of the stellar component, corresponding to the highest part of the bulge. The radius of this ring (at the top of the cylinder) is slightly smaller than the outer radius of the main ring (at the bottom), which makes them appear close but still distinct in face-on projections (e.g. \fig{morpho}). The curved (side) surface of the cylinder is made of ring material with various vertical velocities, and of the initially outermost material which has experienced a weaker perturbation from the companion.

\begin{figure}
\includegraphics{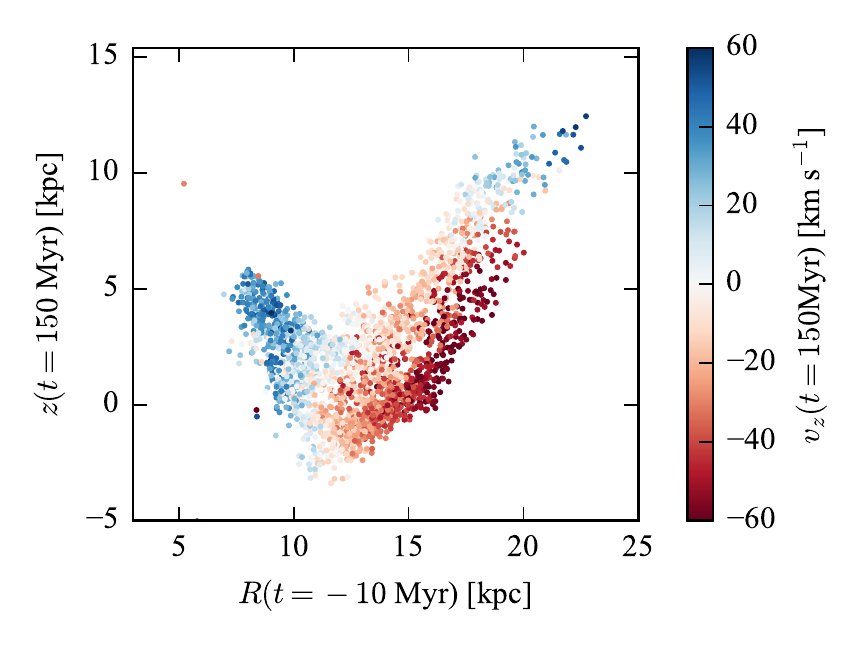}
\caption{Vertical offset of stars with respect to the main plane of the galaxy (measured $150 \Myr$ after the collision), as function of the galactocentric radius (measured $10 \Myr$ before the collision). Colour codes the vertical velocity after the collision. Only a randomly selected subset of the stars is shown, for the sake of clarity.}
\label{fig:cylinder}
\end{figure}

\fig{cylinder} links the vertical distribution within the cylinder to the radial distribution before the collision. The highest material and that with the most positive vertical velocity originate from the outermost disc. Comparably high vertical velocities are also found at lower positions, where lies the material originating from the inner off-nuclear region. The actual offset of the trajectory of the companion with respect to the $z$-axis induces most of the dispersion shown in the distribution of \fig{cylinder} which gets narrower if we limit ourselves to a smaller azimuthal fraction of the cylinder.

Stellar and gaseous densities on the side of the cylinder are $\sim 10\mh 15$ times lower than that of the clumps in the main plane, and as a result this region does not host any star formation activity.  Our simulations do not indicate any strong misalignment between the gas and the stars in this structure, although the gaseous component appears to be slightly more extended (due to its initially lower binding energy to the galaxy), in a similar manner as tidal tails in classical interacting pairs.

We note a mild (and temporary) accumulation of gas at the top edge of the cylinder, due to the distribution of velocities. This translates into the thin ring of gas noted above (recall \fig{cylinderslice}). A comparable configuration is detected in AM~0644-741 (also known as the the Lindsay-Shapley ring galaxy, \citealt{Graham1974}) where a thin ring close to the inner radius of the main ring is clearly visible in neutral gas \citep{Georgakakis2000}, as well as in \ha and infrared \citep{Higdon2011}, indicating a star formation activity (see also \sect{tides}). In our models, the lower densities of the top of the cylinder yield however a lower amount of fragmentation than along the main ring, so that the former ring is smoother and more regular (much more diffuse) than the latter.

The positive vertical velocities are reduced by the gravitational acceleration of the companion and the main plane of the galaxy, before becoming negative. The material in the cylinder lags behind (vertically) the main plane. The cylinder thus flattens, then expands below the main plane, and this oscillating evolution goes on in a damped manner (due to weakening of the effect of the companion galaxy, and the strengthening of that of the main galactic plane).

\subsection{Shear}
\label{sec:shear}

\begin{figure*}
\includegraphics{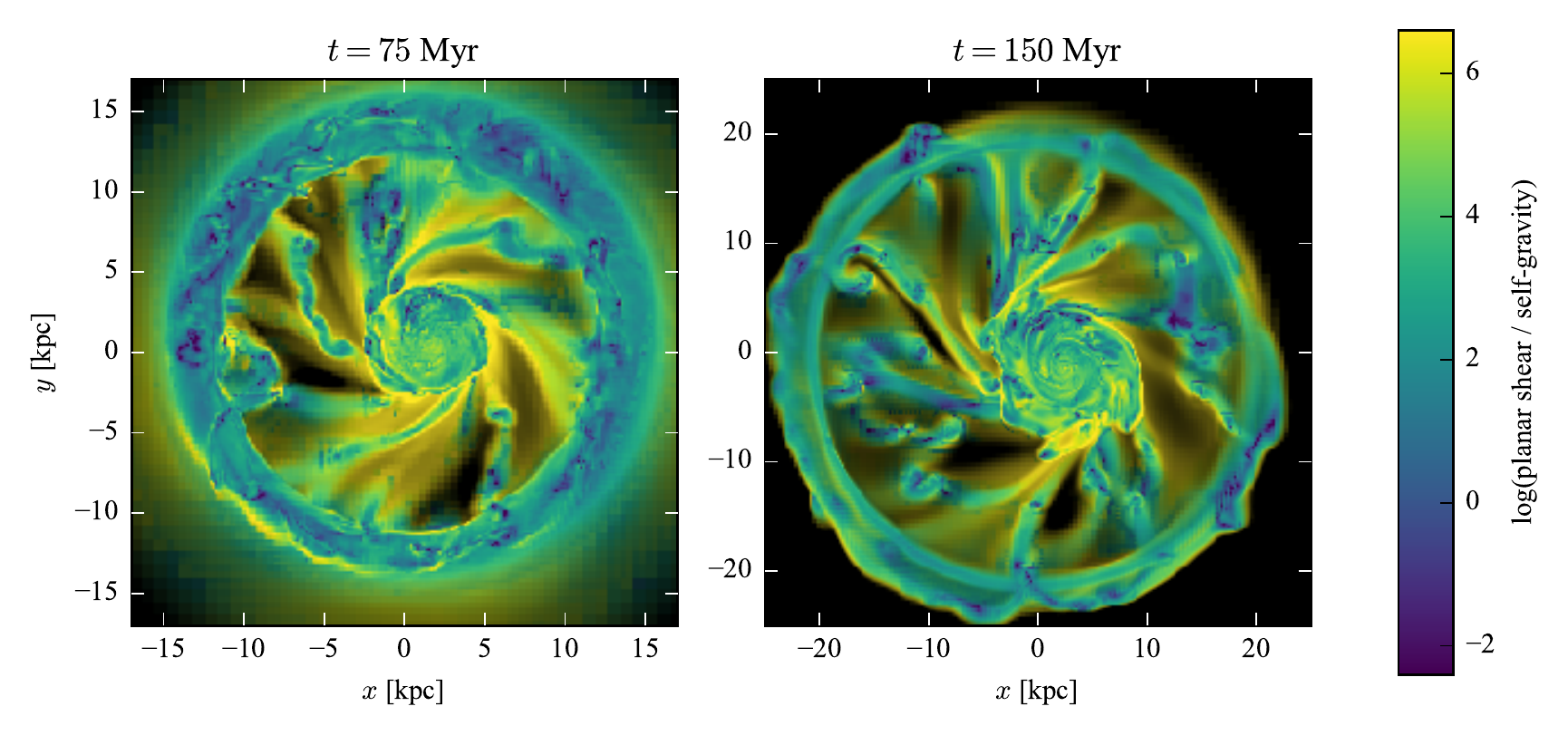}
\caption{Map of the shear in the $x\mh y$ plane, normalised to self-gravity (see text), 75 and $150 \Myr$ after the collision. To highlight the position of the gas structures, the luminosity of this map represents the gas density, such that bright areas correspond to dense gas.}
\label{fig:shearmap}
\end{figure*}

\fig{shearmap} shows the maps of shear in the $\qt=1$ case, $75$ and $150 \Myr$ after the collision. The quantity plotted is the logarithm of
\begin{equation}
\delta_x \left[ \left(\frac{\partial v_x}{\partial y}\right)^2 + \left(\frac{\partial v_y}{\partial x}\right)^2 \right]  \frac{1}{G\delta_x \rho}
\end{equation}
where $\delta_x$ is the scale at which the velocities ($v_x$ and $v_y$) and gas density $\rho$ are measured ($100 \pc$ here). The rightmost fraction represents the inverse of the self-gravitational force and the rest is the planar shear expressed as a centrifugal force. For simplicity, we refer to this quantity as ``shear'' in the following.

As the ring sweeps disc material to radii larger than those before the collision, the ring yields, on average, a higher circular velocity than the material outside, at larger radius. As a result, a strong shear is found along the outer boundary of the ring (particularly visible in the lower-left corner of the left panel of \fig{shearmap}). Inside the ring itself, the material is well mixed resulting in a rather smooth distribution of velocities and thus a zone of low shear. Locally, self-gravitating over-densities yields the minimum levels of shear found in the system.

The spoke features are shaped by the large-scale dynamics (left panel in \fig{shearmap}). Strong shear is found along the edges of the spokes and likely participates in keeping them thin. Because of the radial dependence of the velocity, shear is significantly weaker along the radial spokes than those with a spiral-like configuration. A similar reasoning has been proposed to explain the different morphologies of star forming regions (beads on a string and spurs) in disc galaxies by connecting the spiral pitch angle to the presence of Kelvin-Helmholtz instabilities in high shear volumes \citep{Renaud2014}. Here we also note that ``low pitch angle'' spokes (i.e. spiral-like) host fewer over-densities than their more radial counterparts.

In the nuclear region, the interplay of intrinsic motions and the influence of accreted material makes a complex shear structure. We note however that the average shear is stronger here than in the main ring, likely due to the fact that the rotation curve is naturally steeper in the inner parts than in the outer ones.

When the spokes are well in place (right panel in \fig{shearmap}), the range of shear values narrows. Both the expansion and the depletion of the ring lower its density and thus its self-gravity, which rises the shear quantity plotted. Dense clumps marked by low shear can mainly be identified in the spokes. When they reach the nuclear region however, their signature becomes more difficult to spot.

Such a configuration indicates a preferential gathering of bound gas clumps and survival to external disruption effects, and thus physical conditions favouring star formation, in the main ring and in the radial spokes. More physical criteria must however be fulfilled to lead to an actual star formation activity.

\subsection{Star formation and its regulation}
\label{sec:sf}

\begin{figure}
\includegraphics{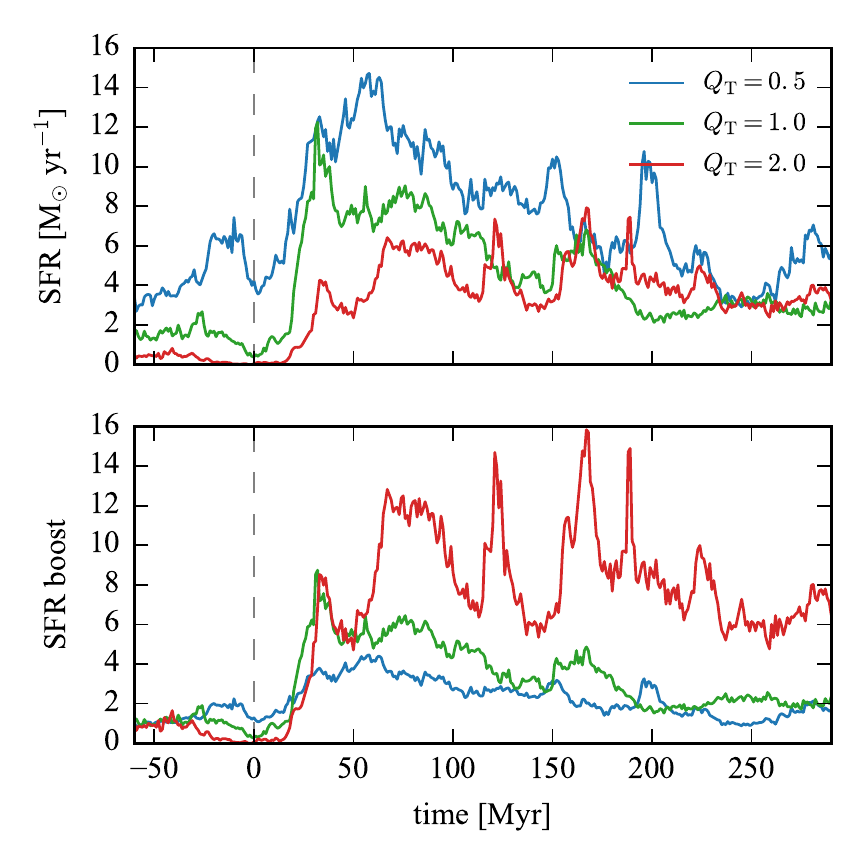}
\caption{Evolution of the SFR in our three simulations (top) and of the SFR normalised to its average value before the collision (bottom) which indicates the enhancement of the star formation activity induced by the collision. The time of the impact is indicated by a vertical dashed line.}
\label{fig:sfr}
\end{figure}

\fig{sfr} shows the evolution of the SFR of the entire system (recall however that our companion is modelled without gas and thus does not host star formation). The intensity of star formation before the collision depends on the overall stability of the disc and thus varies with $\qt$. By normalizing the SFR to its average pre-collision value, the bottom panel of \fig{sfr} shows the relative enhancement triggered by the collision. After a short episode of quenched star formation (see next section), the SFR rises rapidly ($t \approx 15 \Myr$) as the ring expands and gathers gas. Details of the star formation activity depend on the intrinsic stability of the discs and are discussed below.

The boost of SFR is less than the maximum value observed for such a mass ratio and pericentre distance in more classical interactions \citep[e.g.][]{Scudder2012}, but it is however clearly visible, even in the least favourable case of $\qt=0.5$ in which some level of fragmentation of the ISM takes place before the interaction. We note that the maximum value of the SFR is not reached immediately after the collision, but can be found in general $\approx 50 \Myr$ after, and even up to $150 \Myr$ later in the form of short bursts in the $\qt=2$ case (due to accretion of gas clumps into the nuclear region, see \sect{sfrloc}). In all cases, the SFR remains above its intrinsic value for several $100 \Myr$, i.e. even after the dissolution of the ring ($\approx 180\mh 200 \Myr$), but star formation is then limited to the central region, as discussed later. 

\subsubsection{Quenching at collision}
\label{sec:quenching}

\begin{figure}
\includegraphics{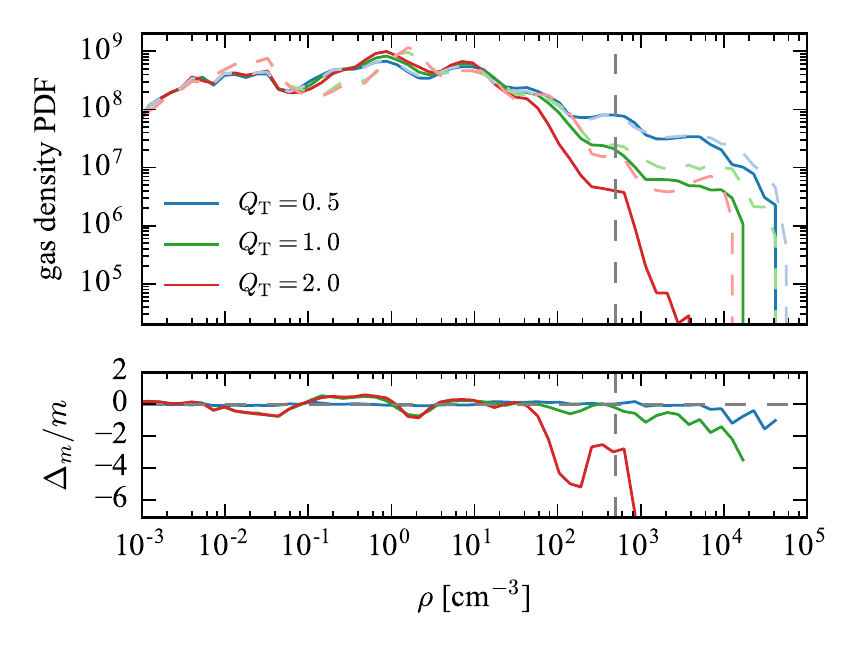}
\caption{Top: mass-weighted gas density PDF of our three models at the moment of the collision (solid lines), and $40 \Myr$ before (light colour, dashed lines). Bottom: relative difference of these PDF between the two epochs. (A negative value indicates a depletion of the gas reservoir at the corresponding density.) The vertical line marks the density threshold for star formation.}
\label{fig:pdfquenching}
\end{figure}

The SFR is reduced in all models a few Myr before the collision, but with an amplitude and a duration depending on the intrinsic stability of the discs. As the companion approaches, 
it induces a strong tidal field, modifying the morphology of the galaxy and disrupting its gaseous clumps. Such disruption smooths out the marginally bound over-densities which thus quenches the star formation activity. Some level of star formation activity is however preserved in the clumps dense enough to survive this tidal disruption. \fig{pdfquenching} illustrates this point by showing the depletion of the dense gas reservoirs as the companion approaches ($-40 \Myr < t < 0$). In the more fragmented case of our $\qt=0.5$ model, the more numerous and denser gas clumps that exist before the collision are able to resist the tidal disruption better than in the other cases ($\qt=1$ and $\qt=2$). Therefore, the dissolution of the dense gas structures and the quenching of star formation occur later and in a less efficient manner than in our other models. In all cases, it takes $\approx 15 \Myr$ after the collision for the interaction-triggered enhancement to overcome its effects.

In any disc galaxy, the gravitational acceleration is weaker along the direction perpendicular to the disc, than in the plane of the disc. Therefore, when such a disc is involved in a collisional ring system with a companion on a highly inclined orbit, the tidal disruption perpendicular to the plane of the target's disc (i.e. vertical in our case) is stronger than in less inclined configurations. Therefore, ring systems are more susceptible to the quenching mechanism detected here than other interacting systems with otherwise comparable mass ratios and/or pericentre distances.

\subsubsection{Schmidt-Kennicutt relation and starburstiness}

\begin{figure}
\includegraphics{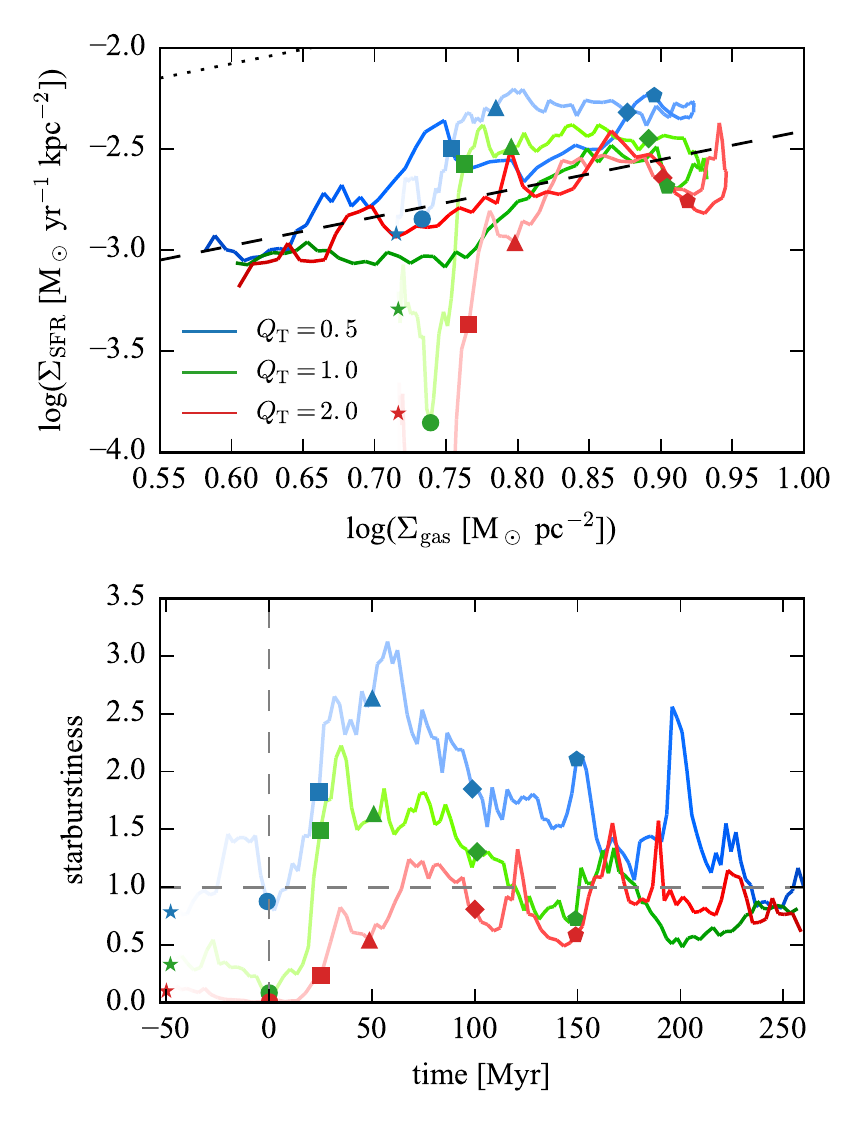}
\caption{Top: evolution of the systems in the Kennicutt-Schmidt plane. The curves darken with increasing time. Symbols mark different epochs (see bottom panel). The dashed and dotted lines show respectively the disc and the starburst sequences as defined in \citet[see also \citealt{Genzel2010}]{Daddi2010}, for reference. Bottom: evolution of the starburstiness parameter (see text). Colours and symbols are the same in both panels.}
\label{fig:ks}
\end{figure}

\fig{ks} shows the evolution of our models in the Kennicutt-Schmidt (KS) plane. The surface densities of all gas ($\Sigma_\mathrm{gas}$) and SFR ($\Sigma_\mathrm{SFR}$) are measured in the face-on view. Only areas denser than $0.1 \Msun/\pc^2$ are considered, and the surface they span is used to compute the two surface densities. We then compute the starburstiness parameter defined as the ratio between the measured $\Sigma_\mathrm{SFR}$ and that of the disc sequence for the corresponding measured $\Sigma_\mathrm{gas}$. A galaxy is considered in a starburst regime when its starburstiness parameter exceeds 4 \citep{Schreiber2015, Renaud2016, Fensch2017}.

All three models have a comparable behaviour. As expected, they start with initially comparable $\Sigma_\mathrm{gas}$ and their $\Sigma_\mathrm{SFR}$ relates to their intrinsic stability. Since we measure the surface densities over the entire disc, the small scale ($\lesssim 1 \kpc$) differences between our models do not appear in $\Sigma_\mathrm{gas}$ (up to slight differences due to gas consumption at different SFRs). Only $\Sigma_\mathrm{SFR}$ is affected by the small scale structures of the ISM.

At the time of the collision, quenching induces a rapid drop of the star formation efficiency (SFE). Immediately after ($0\mh 25 \Myr$), the gas becomes slightly denser, and the SFE rises rapidly. The enhancement of star formation takes over the quenching mechanism. As the ring forms and gathers gas, the gas surface density increases. It reaches its maximum at $\approx 125 \Myr$, i.e. after the ring gas reservoir started to deplete into the spokes (recall \fig{ringmass}). The two surface densities decrease, roughly following the disc sequence ($\Sigma_\mathrm{SFR} \propto \Sigma_\mathrm{gas}^{1.42}$, \citealt{Daddi2010}).

According to our convention, none of our models reaches the starburst regime (starburstiness $> 4$). However, galaxies initially more efficient at forming stars, i.e. with a higher intrinsic starburstiness parameter could temporarily reach the starburst sequence in the KS plane and be classified as luminous infrared galaxies (LIRGs, \citealt{Kennicutt1998b}). However, the non-linearity of star formation and feedback requires to run more simulations to check this particular point.

\subsubsection{Star formation as function of location}
\label{sec:sfrloc}

\begin{figure*}
\includegraphics{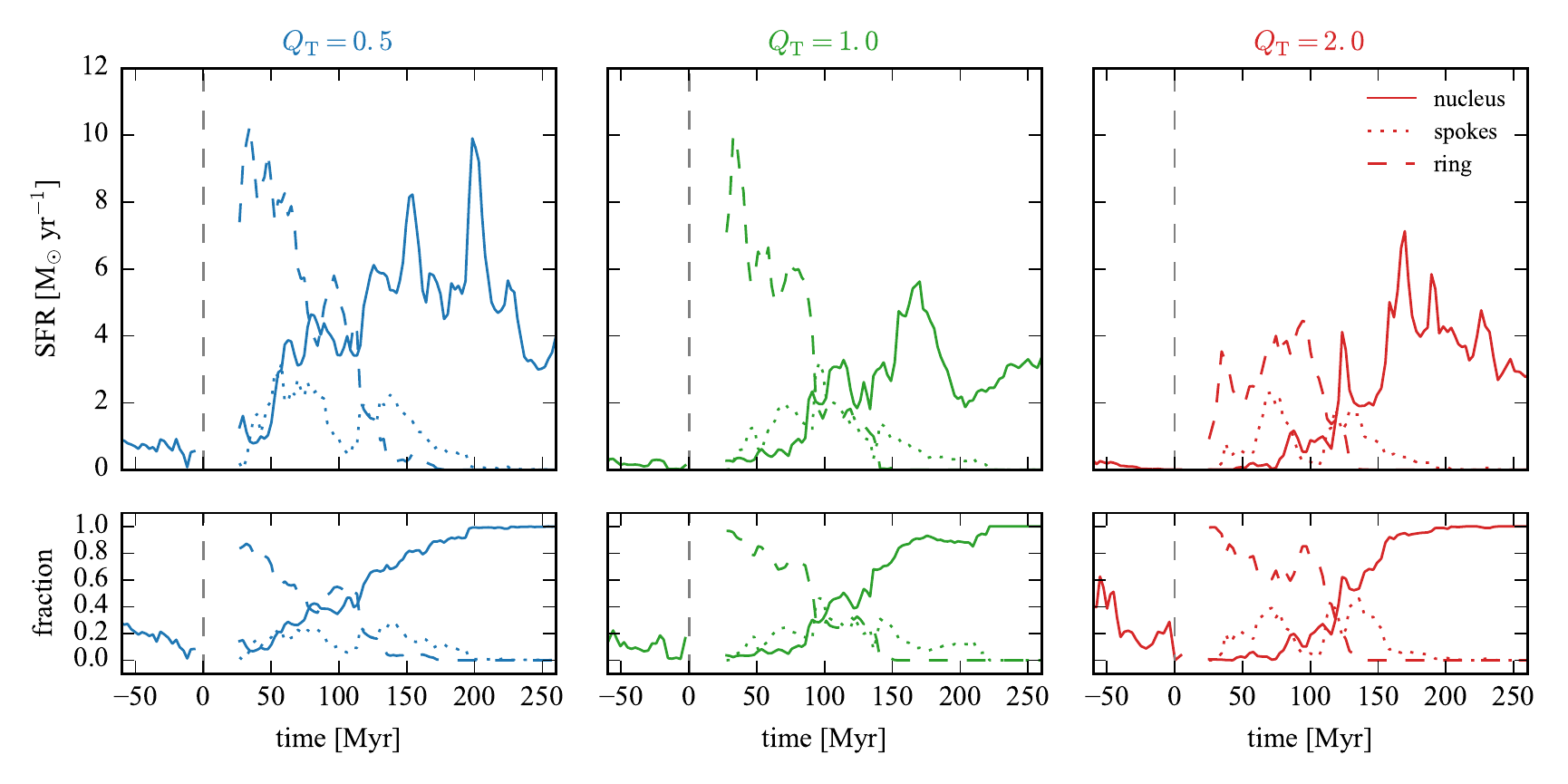}
\caption{Star formation rate in the nuclear region, the spokes and the ring, for our three cases (top) and corresponding fraction of the total SFR (bottom). Data is missing for $0 < t < 25 \Myr$, due to the difficulty of identifying the structures while they form. Before the collision ($t<0$), we assume the nucleus has a constant radius of $1.4 \kpc$.}
\label{fig:offnuclear}
\end{figure*}

\fig{offnuclear} shows the evolution of SFR separately in each of the three main regions of our Cartwheel-like model. Before the collision, we arbitrarily define the radius of nucleus as provided by our parametrisation for $t=0$ (recall \sect{morpho}), which leads to $1.4 \kpc$ and roughly matches the size of the central structures when identified by eye. Once the post-collision structures can be identified ($t > 25 \Myr$), most of the star formation activity occurs in the ring during the first $\approx 100 \Myr$. The ring expands with time, becoming thicker in the $x\mh y$ plane, while some of its gas is fuelled along the spokes toward the nucleus (\sect{morpho}). The ring thus becomes more and more diffuse. As a result, the SFR there decreases while that of the nucleus increases, and this transition is visible in the activity in the spokes. This trend is found for all three models: only the details and the fast-evolving, small-scale features differ. For instance, the starburst in the nucleus is delayed after the encounter by $\approx 50$, 80 and $100 \Myr$ in the $\qt = 0.5$, 1 and 2 simulations, respectively.

\begin{figure}
\includegraphics{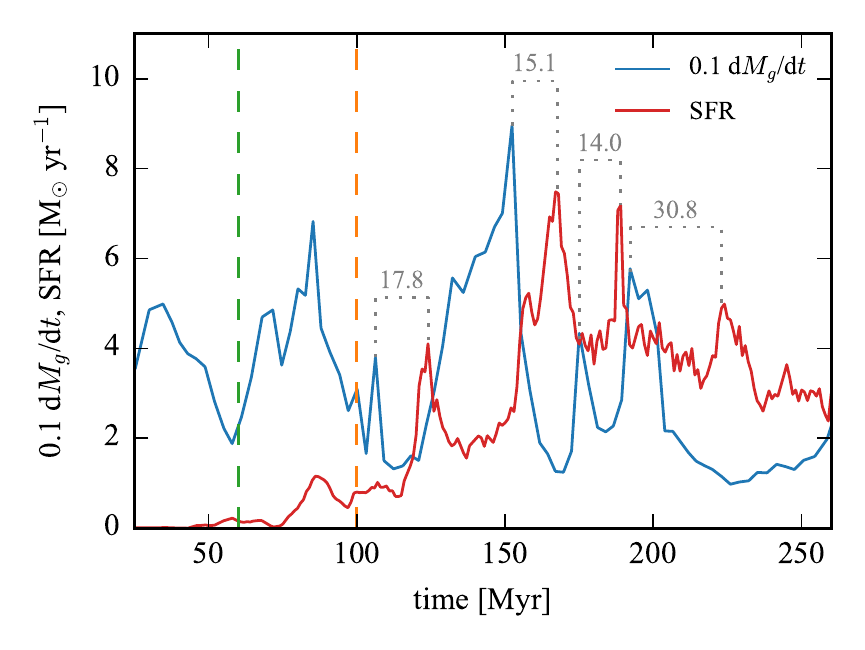}
\includegraphics{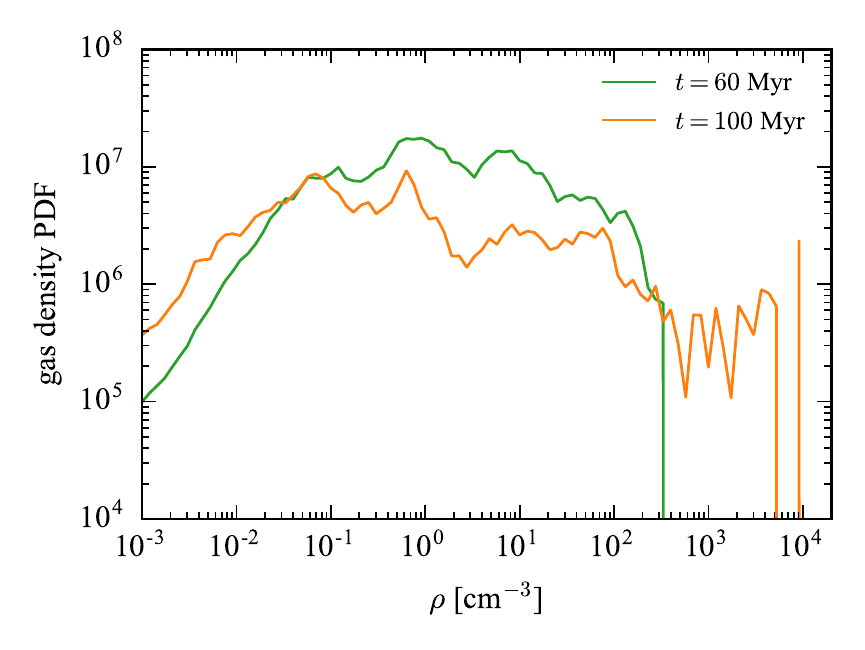}
\caption{Top: evolution of the net gas inflow rate (computed as the time derivative of the mass), and star formation rate in the nuclear region, after the collision, for our $\qt=2$ case. In a first phase ($t \lesssim 70 \Myr$), the accretion of gas does not translate into a significant rise of the SFR. After, peaks in the accretion rate are related with rapid increase of the SFR, with a time delay of $\approx 14\mh 30 \Myr$, as indicated in grey. The vertical green and orange dashed lines mark the epochs when the mass-weighted gas density PDF is computed (bottom). These PDFs are measured in the innermost $2 \kpc$ of the spokes region (i.e. close to the nucleus but still outside it, just before this material gets accreted by the nucleus).}
\label{fig:inflows}
\end{figure}

In particular, the SFR in the nucleus of the case $\qt=2.0$ yields strong bursts, but only at a late stage of the evolution ($t \gtrsim 80 \Myr$, \fig{sfr}). \fig{inflows} shows that such bursts of star formation correspond to gas accretion onto the nuclear region. As shown in the bottom panel of \fig{inflows}, the accreted gas yields a dense component, corresponding to dense gas clumps which trigger a local starburst shortly after being accreted onto the nucleus. We note that a comparable situation is found in the $\qt=0.5$ and $\qt=1$ cases, intrinsically less stable, but with bursts of milder intensity than in the $\qt=2.0$ model, relative to the pre-interaction SFR (recall \fig{sfr}). 
\begin{figure}
\includegraphics{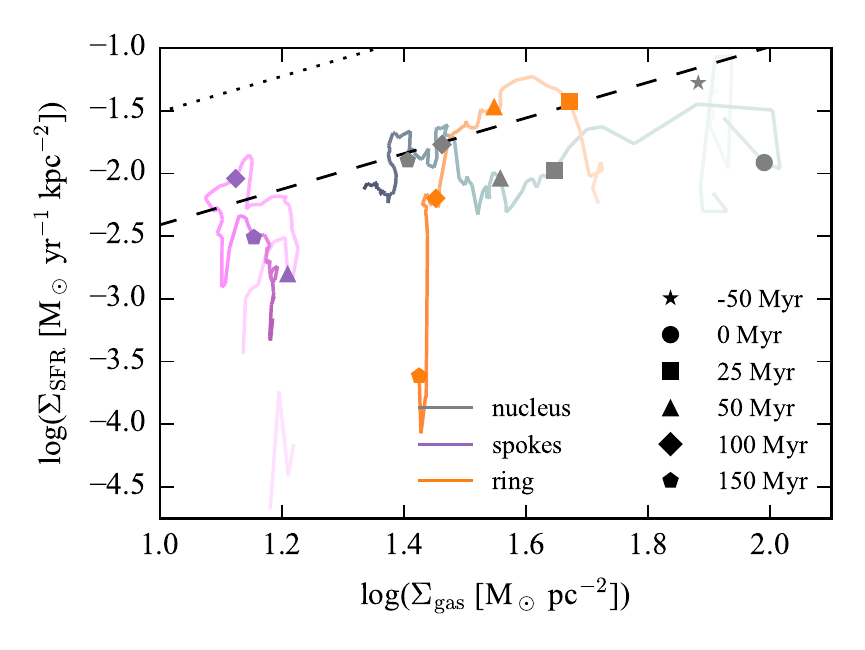}
\caption{Same as \fig{ks} but telling apart the nuclear, spokes and ring regions in our $\qt=1$ model. Note that the ring and the spokes are not defined before $t =0$. The times associated to the symbols are as in \fig{ks} and indicated in the legend.}
\label{fig:ksloc}
\end{figure}

\fig{ksloc} shows the evolution of the sub-regions in the KS diagram, using the same method as before (for the $\qt=1$ model only). When it is first detected, the ring is rather diffuse and hosts very little star formation. In the process of ploughing gas, it rapidly grows in mass and becomes more efficient at forming stars. However, its radial expansion leads to an increase of its surface, resulting in the net decrease of its gas surface density. During the course of the simulated evolution, the ring will host the most efficient star formation activity, at $t \approx 30 \Myr$, because of low shear (\sect{shear}) and other factors like tidal compression (see \sect{tides}). This efficiency then starts to slowly diminish before the ring loses mass (recall \fig{ringmass}) but drops as its reservoir depletes faster than it accretes ($t \approx 100 \Myr$), until all star formation activity ceases.

A comparable but delayed overall evolution is found in the spoke regions. The gas density is significantly lower than in the ring but the star formation efficiency reaches comparable (yet slightly lower) values. The maximum is reached approximately when the mass of the ring is maximal. (Recall the ring starts losing mass through the spokes before.) We note that the gas surface density of the spokes varies little during the evolution ($\approx 25\mh 150 \Myr$), as the combined result of the expansion of ring (which makes the spokes longer) and the accretion of gas from the ring, and its loss to the nucleus.

Before the collision, star formation in the nuclear region is rather stochastic and discontinuous. The tidal disruption from the companion is visible as a lowering of the star formation efficiency. Then follows a rapid decrease of the gas density due to the growth in size of the nucleus and the formation of the secondary, inner ring (creating a gap of low density inside it). This decrease slows down later ($t \approx 100 \mh 150 \Myr$) when the nuclear region is fuelled with gas from the ring (via the spokes), and is thus accompanied with a rise of the star formation efficiency. At the end of the period we explore, star formation only occurs in the nucleus (see also \fig{offnuclear}), with an efficiency comparable to that of the nuclear region in the pre-collision phase.

The star formation activity thus follows a sequential evolution, from the ring to the spokes and then the nuclear region. The observations of Cartwheel, however, do not report any star formation in the central area \citep[e.g.][]{Amram1998, Horellou1998, Charmandaris1999}. This suggests that the real Cartwheel is still in the phase where all star formation occurs in the ring itself (i.e. $t \lesssim 50\mh 100 \Myr$ according to our models, depending on \qt). We note, however, that other considerations based on the morphology of the inner ring and the size of the structures rather support a match to observations by our models at a later epoch \citep[see also][]{Horellou2001}. The discrepancy between this criteria then would question the full representativeness of our model to the real galaxy.

In any case, the general evolution we describe here should be retrieved in most ring galaxies with spokes, and thus the spatial distribution of star forming regions could be used to estimate the evolutionary stage of the system, and date the collision.

\begin{figure}
\includegraphics{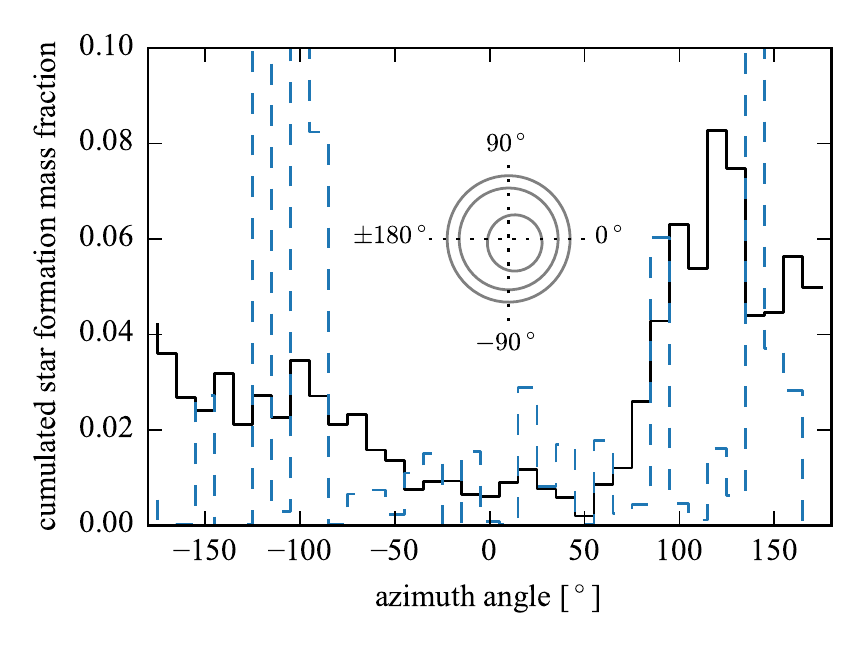}
\caption{Azimuthal distribution of the mass of the stars formed in the ring. In black, this mass is cumulated over the full duration of the ring detectability, i.e. between $t = 25$ and $180 \Myr$. In dashed blue, only the brightest stars (i.e. younger than $30 \Myr$) are accounted for at $t=50 \Myr$, when the star formation activity in the ring is still important and allows us to derived significant azimuthal information. Both distributions are statistically comparable and yield the same first-order asymmetry. The azimuth angle is measured counter-clockwise in the $x\mh y$ plane, as shown in the cartoon representing the configuration of ring and the position of the nuclear region at $t=150 \Myr$.}
\label{fig:azimuth}
\end{figure}

\fig{azimuth} shows the azimuthal distribution of the stellar mass formed in the ring, cumulated over the entire ring lifetime, and when considering young, bright stars only (at $t=50\Myr$). The distribution of young stars only suffers from low number statistics, and mostly traces the formation of a handful of massive star clusters, but overall corresponds to the cumulated distribution discussed here, in particular through the strong deficit of star formation on the right half. In the coordinate system we adopted, a clear left/right asymmetry is found with star formation mainly occurring in the left-hand side regions ($78\%$ for both the young star distribution and the cumulated one). We also note a second-order top/bottom imbalance, slightly favouring the top half ($56\%$ for the young star distribution, which is dominated by a couple of star clusters at $+90^{\circ}$ and $+140^{\circ}$, and $60\%$ for the cumulated distribution). This distribution is related to the position of the nucleus with respect to the ring.

\begin{figure}
\includegraphics{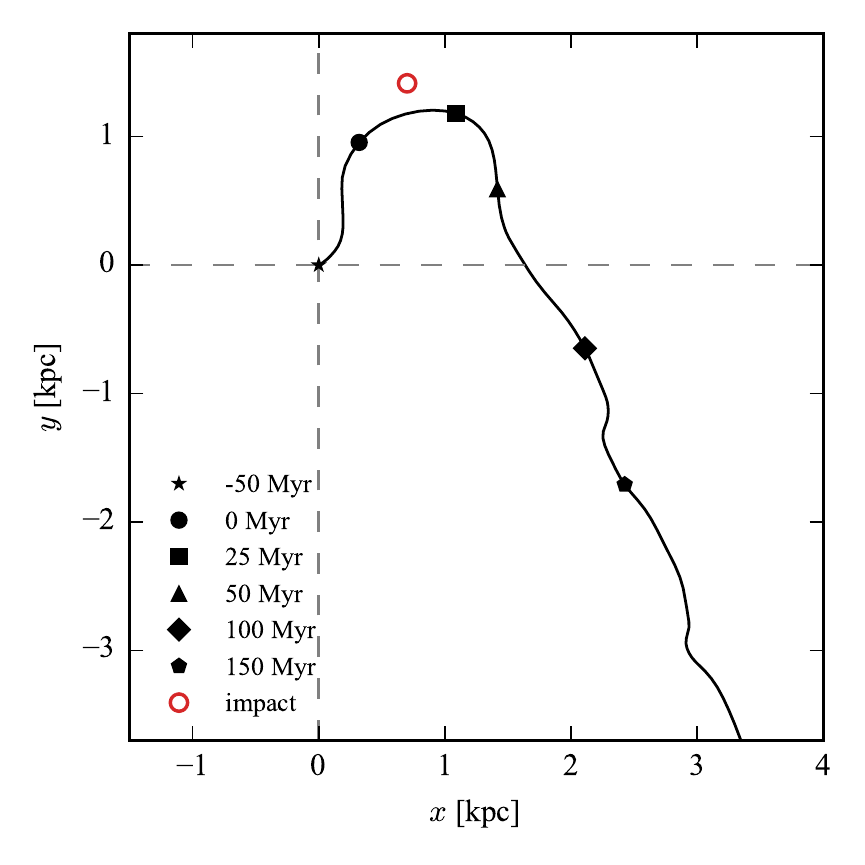}
\caption{Trajectory of the nucleus in the reference frame centred on the ring's centre. The positions at several epochs are marked by symbols, and the centre of mass of the companion galaxy at the time it crosses the mid-plane of the main galaxy (defined as $t=0$) is shown as the red open circle.}
\label{fig:com}
\end{figure}

\fig{com} shows the trajectory of the centre of mass of the most bound particles of the nuclear region, in the reference frame centred on the ring's centre. Because of the off-centred impact point of the interloper galaxy toward the top-right side ($x \approx 380 \pc$, $y \approx 460 \pc$, with respect to the position of the nucleus at $t=0$, i.e. an azimuth of $+50^{\circ}$, see \fig{azimuth}), the nuclear region is first attracted by the companion toward the top-right quadrant, before moving to the bottom-right part. This suggests that the presence of the nucleus on the right-hand side of the system reduces star formation with respect to the other side (i.e. our first-order left/right imbalance). The second order effect (top/bottom) is much weaker and more difficult to analyse as it evolves with time, like the position of the nucleus and the global SFR do. It is thus likely that the net top/bottom distribution results from the interplay of several mechanisms and should not be solely connected to the position of the nucleus. The azimuthal distribution of \fig{azimuth} however strongly suggests that the presence of the nearby nuclear region plays an important role in regulating the star formation activity in the ring. Since the intensity of tidal fields yields a $\propto R^{-3}$ dependence, even small asymmetries in the position of the nucleus like those in Cartwheel can impact the ISM structure and star formation activity in the ring. This point is further explored in the next section.

The asymmetry found in the star formation activity in our models corresponds to that observed in the real Cartwheel galaxy, where the part of the ring the furthest from the nucleus hosts the maximum surface brightness \citep[in bands K and B,][]{Marcum1992} and star formation activity \citep[traced by \ha,][]{Higdon1995, Mayya2005}. The Lindsay-Shapley ring galaxy (AM~0644-741) also displays an enhanced star formation activity in the furthest quadrant of its ring to the nuclei \citep{Higdon2011}. We note however that the ring galaxy NGC~922 shows a stronger star formation activity on the side of the nucleus than on the opposite one \citep[e.g.][]{Wong2006, Meurer2006}, contrarily to our simulations of the Cartwheel. However, the strong asymmetry observed in the distribution of gas and old stars clusters \citep{Pellerin2010, Prestwich2012} indicates that the ``ring'' structure itself is asymmetric (which thus biases the distribution of star forming regions). Furthermore, the ``ring'' feature observed in NGC~922 shows a sharp outer edge but a much weaker contrast on its inner side and thus differs from the morphology of the Cartwheel. Therefore, it is likely that the observed asymmetry of NGC~922 results from the orbital configuration of the interloper galaxy (and maybe, to some extent, from the pre-interaction morphology of the target disc), while that of Cartwheel only appears and develops after the collision.

\subsubsection{Tides}
\label{sec:tides}

The tidal field is the expression of the local curvature of the gravitational potential and thus, it is compressive in all directions in cored configurations (as opposed to cusps, see e.g. \citealt{Renaud2009}). According to Newton's second theorem, outside a mass distribution, where it can be considered spherically symmetric, the potential can be approximated by that of a point-mass (i.e. a cusp). In that case, the induced tidal field is extensive (e.g. as in the case of the Moon's tidal influence on the Earth). When this assumption breaks (i.e. inside the distribution of matter and/or for non spherical systems), the potential \emph{can} be cored, corresponding to compressive tides.

This situation is frequently found in the central region of galaxies and where the potentials of two spatially extended distributions of mass overlap. \citet{Renaud2009} show that a wide range of galactic interactions (if not all) trigger the enhancement of tidal compression, in terms of intensity of the tidal force and of volume spanned, for $\sim 10 \mh 100 \Myr$. \citet{Renaud2014b} further argue that tidal compression is transmitted to turbulence which itself becomes mainly compressive and doing so, generates an excess of dense gas at $\sim 1-100 \pc$ scales which leads to an enhancement of star formation. Whereas this mechanisms is not the only one at play in interactions and mergers \citep[see also][]{Jog1992, Barnes2004}, it is, to date, the only one explaining the spatial distribution of some actively star forming regions \citep[see also][]{Renaud2015}. However, the configurations and the role of compressive tides is yet to be established in the context of ring galaxies, when the overlap of the two galactic potentials is short-lived and unique, as opposed to more classical galaxy mergers which usually experience multiple passages during which the relative velocities of the progenitors are usually smaller. The ring galaxy configuration also differs from fly-by encounters (which otherwise yield comparably high orbital velocities and a unique pericenter passage) by the small impact parameter of the companion.

In this paper, we consider the net tidal effect, i.e. the differential total gravitational acceleration at a given scale. This encompasses the tidal effect from the companion galaxy, but also that of substructures in the main galaxy itself. Tides are then considered as the external effect on a specific region of the galaxy (which participates in the local dynamics of the region), and not only the sole effect of the companion.

\begin{figure}
\includegraphics{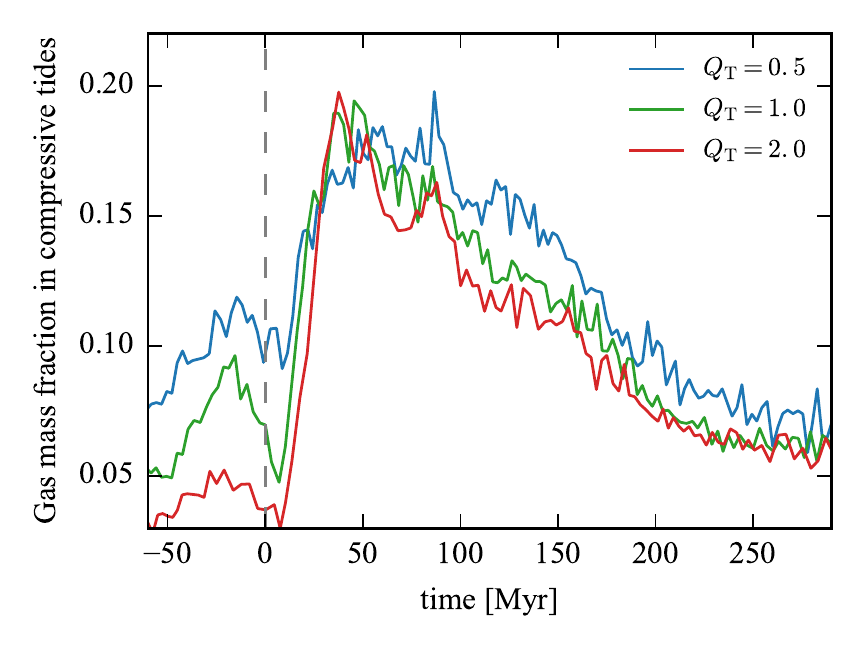}
\caption{Evolution of the gas mass fraction in compressive tides, computed at the scale of $100 \pc$.}
\label{fig:compressivefraction}
\end{figure}

\fig{compressivefraction} shows the evolution of the gas mass fraction in compressive tides.\footnote{The tidal field is evaluated through the tidal tensor computed as in \citet{Renaud2014b}, i.e. with first order finite difference of the gravitational acceleration, but at the scale of $100 \pc$ due to the lower resolution of the simulation presented here. Tides are compressive when all the eigenvalues of the tidal tensor are negative.} The gravitational influence of the gaseous and stellar clumps affects the local tidal field around them (i.e. mainly in dense gas which thus comprises a significant fraction of the gas mass), usually (but not always) making it compressive. Therefore, our least stable galaxy model which comprises many of such clumps yields a larger gas mass fraction of compressive tides already before the interaction than the other cases. 

\begin{figure*}
\includegraphics{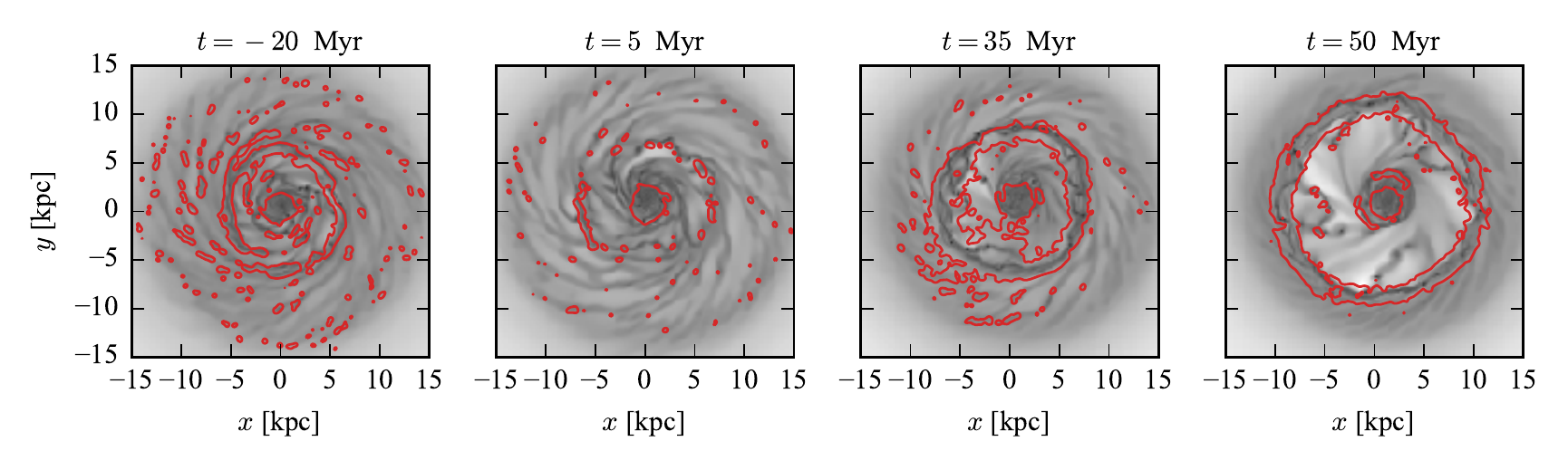}
\caption{Contour maps of compressive tides, overlaid on density maps of the main galaxy seen face-on in our $\qt=1$ model, before the collision ($t=-20 \Myr$), when the mass fraction in compressive tides reaches its minimum ($t=5\Myr$), its maximum ($t=35 \Myr$) and once the ring becomes clearly separated from the nuclear region ($t = 50 \Myr$). The contours have been smoothed using a Gaussian filter with a standard deviation of $200 \pc$, for clarity.}
\label{fig:morphotides}
\end{figure*}

\fig{morphotides} shows the position of the compressive regions at several epochs. A few Myr before the collision, compressive tides are found in the vicinity of star forming regions, i.e. mainly along spiral arms, and in the central volume. As the companion approaches, the disc is perturbed (recall \sect{cylinder}) and extensive tides strengthen across the galaxy. The main tidal force on the target galaxy acts perpendicularly to the plane of the disc, and thus faces less binding resistance than if it was closer to the plane of the disc. Tides thus affect the disc morphology in a more critical manner than in other interactions (recall e.g. \fig{cylinder_sequence}). At short distances, the overlap of the potentials of the two galaxies leads to a tidal field with a compressive contribution, but which is buried under the extensive deformation of the main galaxy. This destructive effect is milder in the central region which retains its intrinsic compressive character, and undergoes a stronger compressive effect from the nearby overlap. However, the importance of compressive tides globally decreases, which translates into the (at least partial) quenching of the star formation activity, discussed in \sect{quenching}. Note that such quenching effect is not observed in interactions with a lower orbital inclination, which better resist the extensive tidal deformation and mostly host compression \citep{Renaud2009}.

At the moment of the collision, the vertical tidal force is minimum and the compressive effect of the overlap takes over. As the companion flies away from the galaxy, the ring forms and expands. The cored nature of its (local) gravitational potential induces tidal compression of a significant amount of the ISM. The global mass fraction in compressive tides rises by a factor of $2\mh 2.5$. This fraction then slowly returns to a lower level as the ring depletes. 

We note that, despite different levels before the interaction, the gas mass fraction tidally compressed is comparable in our three galaxy models after the collision. This relates to the fact that the tidal compression triggered by the galactic interaction spans a volume several times larger than the typical scales of the gas instabilities. Therefore, the $\lesssim 100 \pc$ differences found between our models, because of small scales clumps before the interaction, become of secondary importance in the global gas mass fraction when the kpc-scale compression occurs.

\begin{figure}
\includegraphics{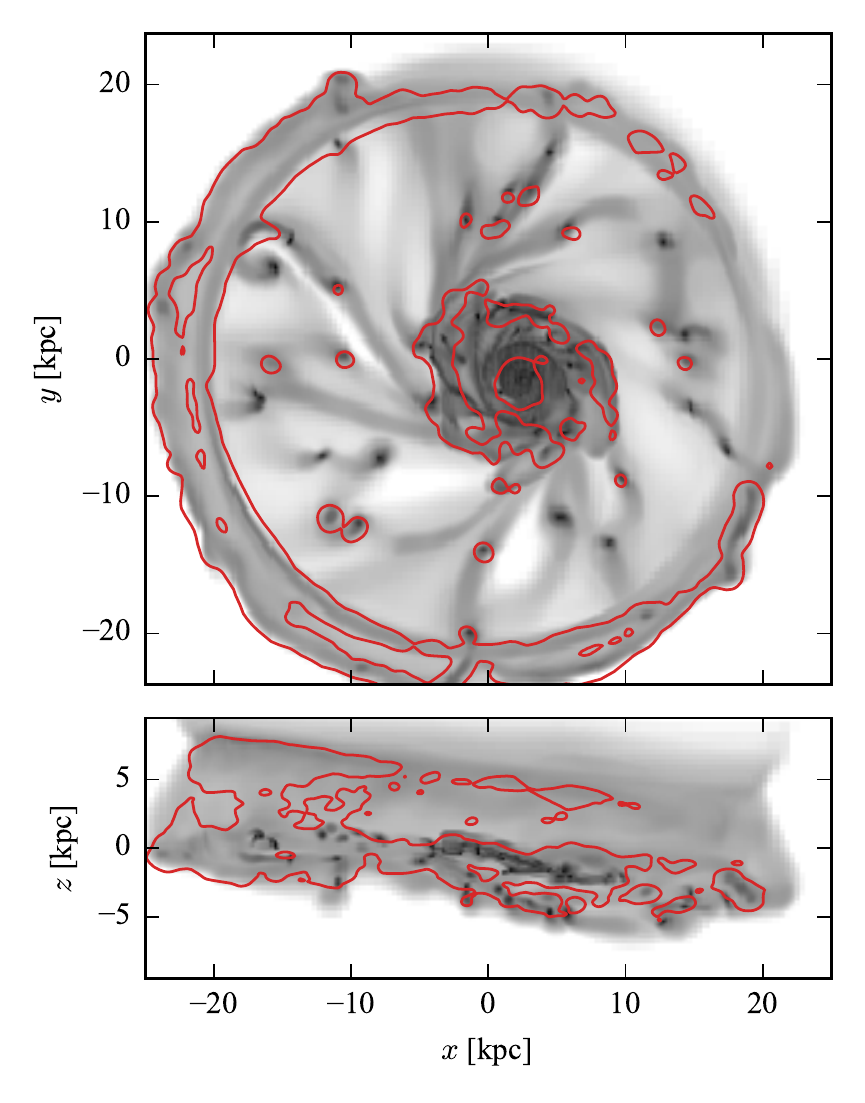}
\caption{Same as \fig{morphotides} but for the face-on (top) and edge-on (bottom) view of the main galaxy, $150 \Myr$ after the collision.}
\label{fig:tidescontours}
\end{figure}

\fig{tidescontours} reveals the position of the tidally compressive regions $150 \Myr$ after the collision (in our $\qt=1$ model, but the others yield comparable large-scale features). The central $\approx 2 \kpc$ of the nucleus are compressive, as even before the collision. While going to larger galactocentric radii, one first finds an extensive region ($2 \kpc \lesssim R \lesssim 4 \kpc$), followed by a compressive secondary, inner ring about $2 \kpc$ wide. 

Outside the nuclear region, most of the volume in the spoke region experiences extensive tides. In particular, the spokes themselves do not yield particularly compressive volumes and do not contrast from the inter-spokes areas. However, several over-densities and gas clumps (found along the spokes) lie in compressive tidal regime.

Along the main ring, we note a strong azimuthal asymmetry: most of the ring is compressive, but the $\approx 30^{\circ}$ wide sector closest to the nucleus (and to the impact point of the companion galaxy) does not host compressive tides. The reason for this feature is the tidal influence of the distant nucleus which is de facto stronger on the closest side of the ring. We note that this absence of compressive tides in this area (and the comparable asymmetry in the intensity of the compression) corresponds to the above-mentioned lower star formation activity along the ring (recall \sect{sfrloc} and \fig{azimuth}). This further highlights the role of the kpc-scale dynamics, via tides, in setting the star formation activity, even $100 \Myr$ after the galactic encounter itself. We therefore speculate that a less massive nucleus (e.g. from a progenitor with a smaller bulge mass) would (i) favour the accumulation of more material in the ring and (ii) have a weaker limiting effect on star formation in the ring, leading to a more evenly distributed star formation.

Furthermore, the dark matter halo (also affected by the gravitational acceleration from the companion) follows the motion of the nucleus and is roughly centred on it (with an offset of only $\approx 300 \pc$ in the $x-y$ plane at $t=150 \Myr$), i.e. is off-centred with respect to the ring. As noted by \citet{Pardy2016}, interactions offset the main components of galaxies, which implies a variation of the local e.g. halo to disc ratio. On top of the variation induced by the tidal field, this could also lead to a change of the local stability of the ISM, which in our case could participate in the azimuthal dependence of the star formation activity along the ring.

\fig{tidescontours} further shows that the over-densities at the top of the cylinder visible in the edge-on view, but also as the thin smooth over-density ring close to the inner radius of the main ring in the face-on projection (recall \fig{cylinder}), are also tidally compressed. Such distribution of densities (over-densities at the top, and in the main plane but hardly any in between) and this configuration of compressive tides is similar to that found close to the the tip of tidal tails developing in more classical galactic interactions where the gas accumulates and forms either tidal dwarf galaxies, or aggregations of stellar structures (see \citealt{Renaud2015}).

In our Cartwheel-like models, the quantity of gas in the cylinder is too small to make this region star forming, but we speculate that other models and/or other collision configurations could lead to star formation there, due to the presence of compressive tides, and the absence of strong shear (\fig{shearmap}) that would likely exist in those slightly different configurations. For instance, a disc with an intrinsic higher gas surface density in the outer regions could lead to the presence of denser gas at the top edge of the cylinder. With the help of the compressive tides, such gas could form star clusters or even tidal dwarfs. In such case, young stars could be detected in two separate planes: the main plane, remnant of the original disc, and the top of the cylinder, several kpc above. Note that the absence of tidal compression in between those planes and the divergent velocity field (\fig{cylinder}) make it unlikely to detect star formation in these intermediate regions, in again a similar way as the bulk of classical tidal tails (except near their extremities where tidal dwarf galaxy can be found). If our interpretation of the features observed in AM~0644-741 corresponding to a vertical cylinder structure is correct (recall \sect{cylinder}), the star formation activity detected there \citep{Higdon2011} would then support the idea presented above.

\subsubsection{Turbulence and compression sequence}
\label{sec:turbulence}

Turbulence is known to be a major actor in setting the density structure of the ISM and thus regulating star formation \citep[see a review in][]{Hennebelle2012}. Its short dissipation time-scale ($\sim 1 \Myr$ for surpersonic turbulence at $\sim 1 \pc$, \citealt{MacLow2004}) implies the existence of (at least) a source maintaining turbulence in clouds. At small scales, stellar feedback (mainly from \hii regions, \citealt{Matzner2002}) contributes to sustaining turbulence in star forming clouds. \citet{Grisdale2017} confirm this point at galactic scale by comparing turbulence energy spectra in simulations with and without feedback. In addition to feedback, turbulence originates from scales larger than the clouds themselves \citep{Ossenkopf2002}, from the infall of gas \citep{Klessen2010}, galactic dynamics (spiral shocks, shear, \citealt{Bournaud2010b, Renaud2013b}), gas accretion \citep{Krumholz2016} and galactic interactions \citep{Renaud2014b}. In a cloud, the relative contributions of the internal and external driving of turbulence thus seem to strongly depend on the galactic environment, and obviously whether the cloud is forming stars or not.

Due to the enormous range of scales involved and because of several severe artefacts introduced by the resolution limits and the structure of the mesh used in simulations, it is necessary to measure turbulence over many resolution elements. In the case of galaxy simulations, this forbids exploring the nature of turbulence at the scale of star formation ($< 1 \pc$). Therefore, we focus here on larger scales to identify variations of turbulence along the interaction, and connect them to other physical processes.

Turbulence can be decomposed into a divergence-free, incompressible mode called solenoidal, and a curl-free compressive mode \citep{Federrath2008}. The solenoidal mode can be seen as the stirring effect of turbulence which tends to smooth out over-densities, while the compressive mode generates excesses of dense gas and increases the density contrast of the medium. When energy equipartition is reached, the solenoidal mode carries $2/3$ of the turbulent energy, while the compressive mode carries the remaining $1/3$.  The distribution of densities thus depends on the nature of turbulence \citep{Federrath2008}. A strong classical (i.e. solenoidal-dominated) turbulence (high Mach number) implies a wide but unimodal density PDF leading to a high SFR, while a compression-dominated turbulence generates an excess of dense gas, explaining the super star formation efficiency of galaxies in the starburst sequence in the KS diagram \citep{Renaud2012}.

\begin{figure}
\includegraphics{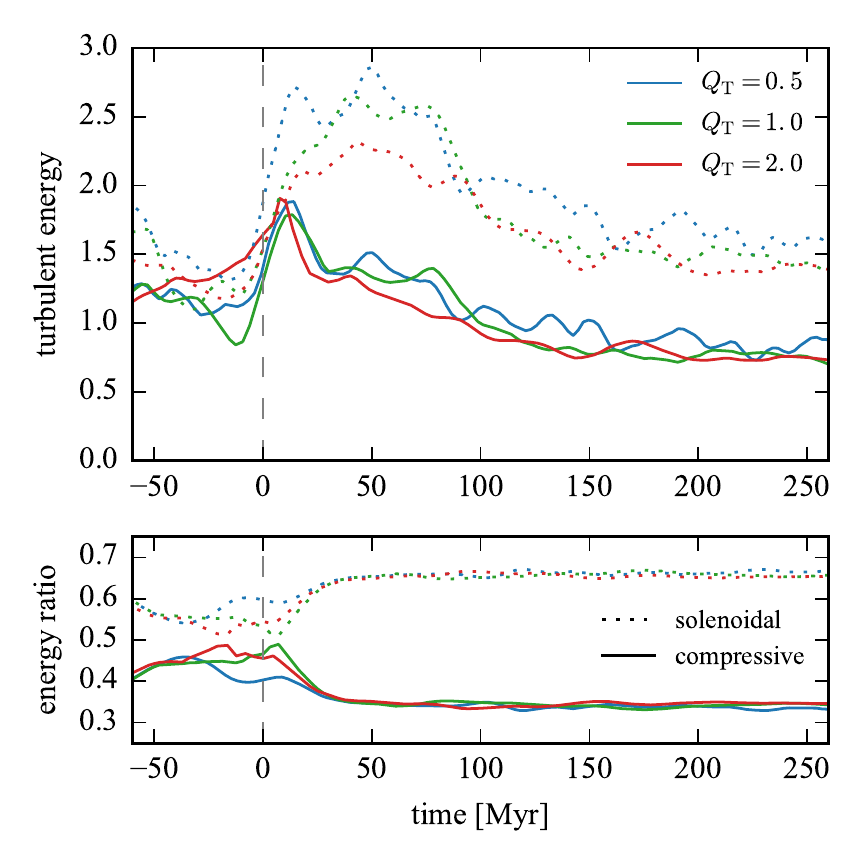}
\caption{Evolution of the energy (top) of the solenoidal (dashed) and compressive (solid) modes of turbulence, and the ratios of the two modes to the total turbulent energy (bottom). Energy equipartition between the two modes would correspond to 2/3 for the solenoidal component and 1/3 for the compressive one (see text).}
\label{fig:turbratio}
\end{figure}

\begin{figure}
\includegraphics{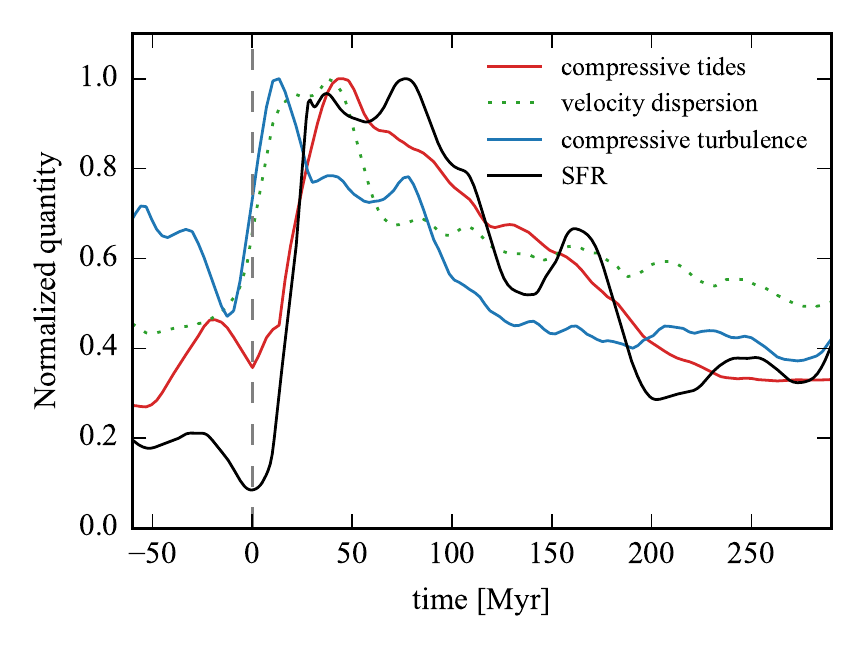}
\caption{Evolution of the main quantities involved in the enhancement of the star formation activity (in our $\qt=1$ case but all models show qualitatively comparable results): gas mass fraction in compressive tides, compressive turbulence energy, velocity dispersion and SFR. To compare their behaviour as function of time, all quantities are normalized to their maximum value, and the curves are smoothed for clarity.}
\label{fig:compressivetrigger}
\end{figure}

We discriminate the two modes of turbulence following the method described in \citet{Renaud2014b}, at the scale of $100 \pc$. \fig{turbratio} shows the evolution of the kinetic energy carried by the compressive and solenoidal modes, and their ratio. Before the interaction, the two modes are (almost) at equipartition. As the companion approaches and the importance of compressive tides rises ($t\approx -20\Myr$, recall \fig{compressivefraction}), the energy in both turbulent modes increases, corresponding to a increase of the velocity dispersion (and the Mach number), as observed in interacting systems \citep{Irwin1994, Elmegreen1995b}. However, this increase does not occur at a constant energy ratio between the two modes. Instead, the compressive mode becomes more important, up to almost taking over the solenoidal component at the epoch of the collision. We note that this effect is milder in the $\qt=0.5$ case than the other, corresponding to the also weaker enhancement of the mass fraction in compressive tides at the same epoch. The energy ratio stalls during the collision, until the companion leaves the target galaxy, and the two turbulent modes resume energy equipartition. The total turbulent energy then slowly decreases, in a comparable manner as the mass fraction in compressive tides and the SFR, as shown in \fig{compressivetrigger}.

The first quantity to increase with respect to its value before the interaction is the gas mass fraction in compressive tides ($t \approx -40 \Myr$). The disruption in the form of extensive tides at collision introduces a short interruption of this raise at $t\approx 0$, as discussed in \sect{tides}. The gravitational, remote effect is then transmitted to the hydrodynamics and alters the turbulence field of the galaxy: the kinetic energy carried by compressive turbulence starts to increase $\approx 35 \Myr$ after the tidal compression (at $t \approx -15 \Myr$). We note that, as for the tides, this increase of the compressive turbulence energy is rapidly stopped ($t\approx 10 \Myr$). However, the velocity dispersion, which encompasses both turbulent modes and is thus equally affected by compressive and extensive tides, retains a high value for a longer period ($t \approx -10 \mh 70\Myr$) than the compressive turbulence energy ($t \approx -10 \mh 20\Myr$). Finally, the enhanced turbulence cascades down to the parsec scales, generates an excess of dense gas (\fig{pdf}) and triggers the starburst episode. We emphasize that the rise of the tidal and turbulent triggers before the star formation activity itself (\fig{compressivetrigger}) demonstrates that the whole process (and particularly the increase of turbulence) is not the sole result of stellar feedback: the contribution of feedback can only appear after the burst of star formation. At later times ($t > 70 \Myr$), the SFR is affected by a number of local phenomena, such as the inflow of gas along the spokes onto the nucleus (\sect{sfrloc}), and its behaviour is less dominated by the large scale effects listed here.

With the notable exception of the strong quenching event, this sequence corresponds to that noted in a model of the Antennae galaxies \citep{Renaud2014b}. However, the changes measured here (mass fraction in compressive tides, velocity dispersion, compressive turbulence energy, widening of the PDF, SFR) are significantly milder than their equivalent found at the pericenter passages in the Antennae. It is likely that the differences originate from the pecularity of the orbit of Cartwheel system and the lower mass ratio than for the equal-mass Antennae, introducing strong extensive tides before the collision (recall \sect{tides}), and limiting the importance of the compressive tidal and turbulent modes. The net effect is a smaller enhancement of the SFR in Cartwheel than in the Antennae (respectively by factors of $\approx 8$ and $\approx 60\mh 80$) shortly after the moments of the pericenters or collisions (\sect{sf}). In both interaction configurations, the long-term enhanced star formation activity (lasting $\sim 100 \Myr$) is to be connected to their intrinsic gas contents, the morphology of the galaxy remnants which harbour nuclear inflows, shocks (e.g. in the ring), and infalling debris along spokes in Cartwheel (\fig{inflows}) and tidal tails in the Antennae. A more detailed exploration of the relative role of the physical mechanisms at play in the enhancement of the SFR in interacting galaxies will be presented in a forthcoming contribution (Renaud et al., in preparation).

\subsection{Young massive clusters}
\label{sec:sc}

Starbursting interacting galaxies host the extreme physical conditions necessary to form massive star clusters \citep[$\gtrsim 10^{5}\Msun$, see e.g.][]{Whitmore1995, ELmegreen1997, Bastian2009}. In the Cartwheel in particular, the detection of ultra-luminous X-ray sources have been repported along the ring, and associated with the presence of young massive clusters \citep{Gao2003}.

Similarities in some properties of young massive clusters in starbursting galaxies and the (supposed) ones of globular clusters at the moment of their formation (redshift $\gtrsim 5 \mh 6$, \citealt{Diemand2005}) led several authors to suggest a link between the origins of the two populations, in low-redshift galaxy mergers and at high redshift \citep[see][and references therein]{Portegies2010}. However, notable differences in their metallicities, the presence of multiple stellar populations and their overall evolution question that young massive clusters would be comparable in present-day globulars if evolved for $\sim 10 \Gyr$ \citep[see e.g.][]{Gieles2016, Lardo2017}. Therefore, it is necessary to establish which conditions and mechanisms (at galactic and cloud scales) lead to the formation of massive clusters, and whether comparable setups can be found in the early Universe. The difficulty of this task lies in the multi-physics and multi-scale nature of the problem. Galaxy simulations reaching parsec-scale resolutions are not sufficiently accurate to properly capture the physics of the individual star forming clouds and introduce many uncertainties in the estimates of the properties of the young clusters \citep[see][]{Renaud2015}. Furthermore, tracking the evolution of these clusters requires capturing their internal physics (stellar evolution, two-body relaxation, e.g. \citealt{Heggie2003}) which is still out of reach of present-day studies over entire galaxies.

In particular, one should keep in mind that the mass and length resolution of our models limit us to the detection of the most massive clusters ($> 10^5 \Msun$, as lower mass objects cannot be distinguished from our stellar particles), and that the softening of the gravitational potential ($6 \pc$) introduces over-estimates in the size (and thus the binding energy) of these systems. However, by limiting ourselves to some properties and focussing on relative comparisons between our models and at different epochs, we can still address several questions on the population of young massive star clusters in Cartwheel.

We identify the star clusters as over-densities in the young stellar population (i.e. of the stars formed after the beginning of the simulation) by running the friend-of-friend algorithm \hop \citep{Eisenstein1998} with the peak, saddle and outer densities of $10^8, 10^8$ and $10^6 \Msun\kpc^{-3}$ respectively, chosen to agree with a visual examination of stellar structures. Then, only stellar particles below the escape velocity of their clusters (computed assuming a point-mass potential) are kept as cluster members. Note that over-densities of old stars (i.e. set up in the initial condition) are also formed during the interaction, but are not accounted for here due to the lack of resolution of this component.

\begin{figure}
\includegraphics{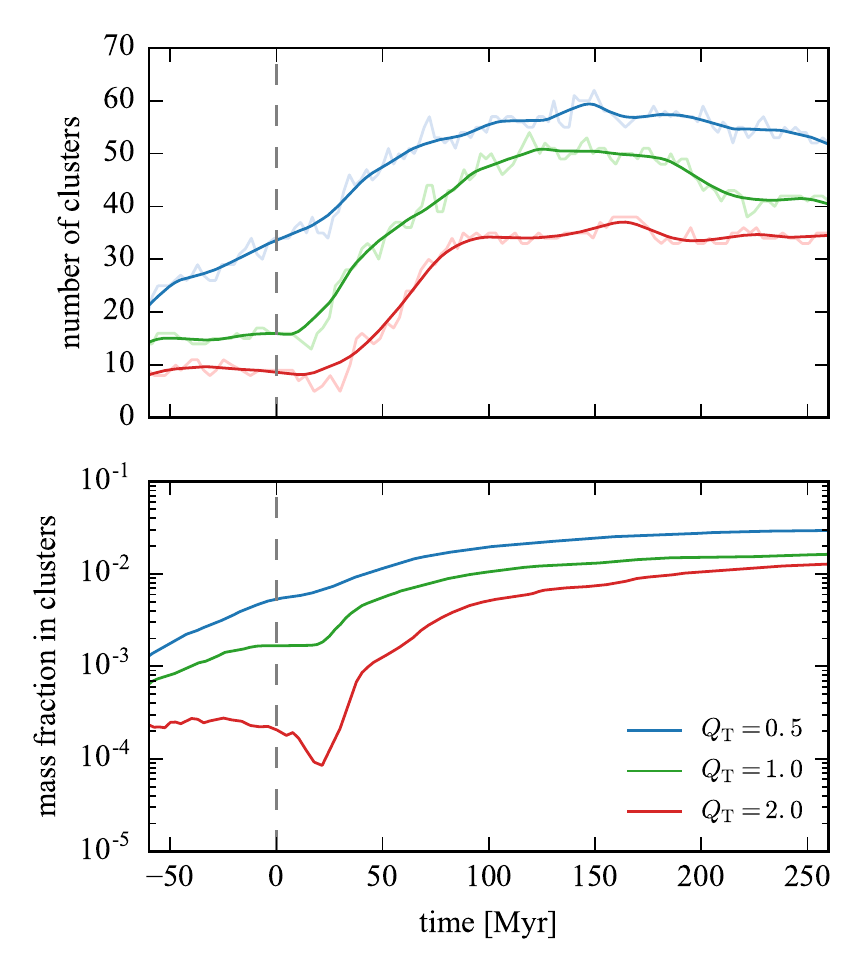}
\caption{Top: number of star clusters identified in our three simulations. The original curves (light colours) have been smoothed out to eliminate the fluctuations introduced by our cluster detection method. Bottom: stellar mass fraction in these clusters as function of time. }
\label{fig:nclusters}
\end{figure}

\fig{nclusters} shows the evolution of the number of star clusters we identify. Uncertainties in our friend-of-friend detection method, specially for the clusters made of a small number of stellar particles (i.e. $10^{4\mh 5} \Msun$) introduce small fluctuations in the total number of clusters identified, from one snapshot to the next. However, the models yield a rapid increase of the number of massive clusters a few Myr after the collision. This increase is however mild in the $\qt =0.5$ which rather shows a constantly rising number of clusters and for which the effect of the collision is present but less visible. We note that the epoch of this increase varies from one model to the next: our $\qt=2$ case is the last one to experience it, about $25 \Myr$ after the collision. This is due to its sensitivity to the quenching mechanism (\sect{quenching}) which has the most severe effect on the SFR for our most stable model. Its less dense clouds being easily tidally disrupted and even destroyed by the approaching companion, it takes this galaxy longer to form new over-densities able to lead to the formation of massive clusters, while the other models host more resistant clouds and yield physical conditions which can more easily and rapidly (re-)form the necessary clumps. 

The number of clusters stalls $\approx 80\mh 110 \Myr$ after the collision, when the mass fraction in compressive tides and the energy of compressive turbulence decrease (\fig[s]{compressivefraction} and \ref{fig:turbratio}). This epoch also corresponds to the moment when the ring starts losing mass (\fig{ringmass}). The number of clusters eventually diminishes a bit, as clusters get dissolved by the galactic tides (e.g. when approaching the nuclear region, or encountering dense structures like the inner ring along their orbit, see \citealt{Gieles2007} for a comparable effect).

The stellar mass fraction found in clusters follows the overall same behaviour, with the notable exception that clusters in our $\qt =2$ model lose mass around the collision, due to tidal stripping from the companion galaxy. Contrarily to the gas clouds, the clusters mainly lose mass after the collision because, being denser, they better resist to tidal harassment, and it takes longer to their stars to reach the necessary energy to escape. We stress, however, that the softening of the gravitational potential in our simulations implies that the cluster binding energy and mass loss rate are biased \citep[see][their figure 7]{Renaud2015}, so that these measurements should only be interpreted qualitatively.

Not enough clusters are formed in this system to derive a statistically significant cluster mass function. However, one can examine the evolution of the mass of the most massive cluster in the simulation. In this analysis, we exclude the central-most object (nuclear star cluster) which is build hierarchically by accreting stars, gas (and forming stars in situ) and star clusters along the simulation \citep[see][and references therein for detailed scenarios of nuclear cluster formation]{Tremaine1975, Milosavljevic2004, Antonini2013, Guillard2016}. In our simulations of the Cartwheel, the most massive cluster before the collision has a mass of the order of $\sim 10^6 \Msun$. That after the collision (i.e. when the SFR is enhanced) is 20, 30 or 40 time larger, for the $\qt=0.5$, 1 and 2 cases respectively. In the case of the modelled Antennae galaxies, this factor reaches 15 at the first pericentre passage. In the Antennae however, the number of clusters jumps by a factor of $\approx 10$, while this number is only $\approx 3$ in our models of the Cartwheel. The resolutions of the simulations are different ($6 \pc$ here and $1.5 \pc$ in the Antennae), but we base our analysis on the \emph{relative} evolution of these quantities before and after the encounters, to limit the potential effect of such discrepancy on our results. Therefore, we can conclude that the change of physical conditions due to the encounter in the Cartwheel favours a more concentrated formation than in the Antennae, i.e. in the form of less numerous but more massive clusters (with respect to the pre-interaction epoch). At first order, and while keeping in mind the limitations mentioned above, the high-mass end of the initial cluster mass function (ICMF) of the Cartwheel would thus be shallower than that of the Antennae. \citet{Kravtsov2005} proposes that the high-mass end slope of the ICMF derives from the shape of the gas density PDF. The relative milder widening of the PDF and the smaller importance of compressive turbulence in Cartwheel than in the Antennae could then explain the differences in the change of the ICMF between the two systems during their respective galactic encounters. This point will be further explored in a forthcoming paper (Renaud et al., in preparation).

\begin{figure}
\includegraphics{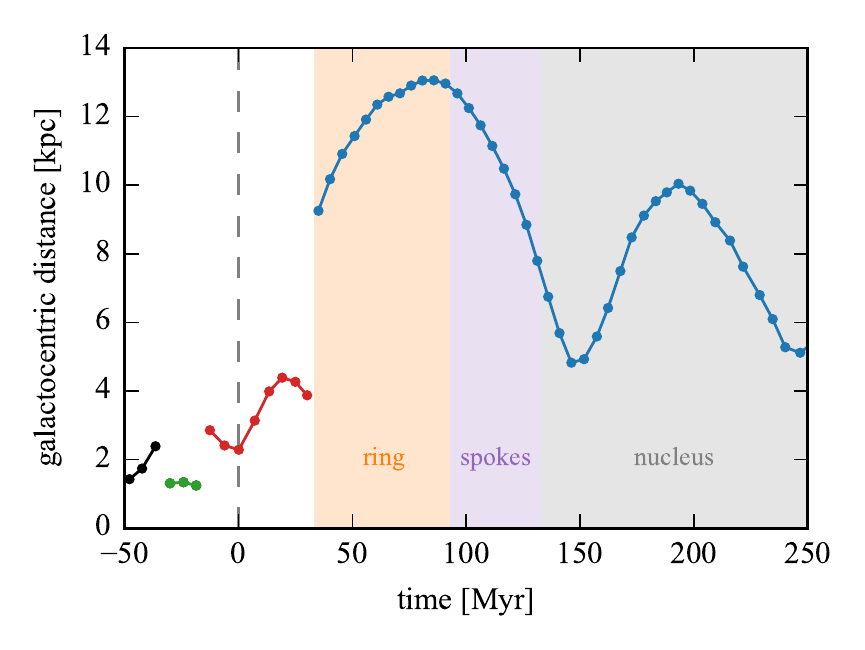}
\caption{Distance to the centre of the nucleus of the most massive cluster in our $\qt=1$ model. Solid lines identify one cluster tracked in time. Discontinuities mark changes in the identity of the most massive cluster. Shaded areas indicate the region in which the most massive cluster is found.}
\label{fig:mostmassive}
\end{figure}

\fig{mostmassive} shows the galactocentric distance (with respect to the centre of the nucleus, i.e. not with respect to the centre of the ring) of the most massive cluster. The rather continuous formation of clusters before the interaction induces several changes in the identity of cluster being the most massive, visible as discontinuities in this figure. They are, however, all found in the inner parts of the disc before the interaction phase ($\lesssim 3 \kpc$, and $\lesssim -10 \Myr$), where the gas density is the highest. A few Myr around the collision ($-10 \Myr \lesssim t \lesssim 35 \Myr$), a cluster, formed in the tidally compressive annular structure visible in the leftmost panel of \fig{morphotides}, remains the most massive throughout the interaction, likely because no new clusters are formed during this time range due to the quenching effect of the companion (\fig[s]{sfr} and \ref{fig:nclusters}). Later, as the ring has formed and is expanding, it encompasses almost all the star formation activity and gathers the physical conditions necessary to form massive star clusters, (at least partly) because of the tidal compression (\fig{morphotides}) and low shear (\fig{shearmap}), resulting in the high star formation efficiency we measure (\fig{ksloc}). A new cluster then takes the role of most massive one in the galaxy and will keep this role until the end of our simulation. It forms in the ring $\approx 35 \Myr$ after the collision, i.e. after the rise of the overall number of clusters formed, but roughly at the time the SFR reaches its maximum value. This indicates that the delay between the collision and the enhancement of star (cluster) formation is longer for more massive objects. This most massive cluster remains in the expanding ring for $\approx 55 \Myr$ and then travels along the spokes to reach the nuclear region $\approx 100\Myr$ after its formation. It does not reach the very centre of the nucleus, but rather remains in the inner, secondary ring. It then moves to larger radii because of expansion of the inner ring, and comes back again to the central region, in a similar manner but slower, as it did from the main ring $\approx 100 \Myr$ earlier. It is however difficult to conclude on the presence of spokes inside the inner ring that could play a comparable role to those discussed above, so that the mechanism driving the cluster toward the centre of the galaxy at late time ($t \gtrsim 200 \Myr$) could well be different.

Assuming that this evolution is common to most ring galaxies (at least those with spokes), the location of the most massive (or brightest) cluster could be used to estimate the epoch of the peak of star formation and constrain the star formation history of the system, while the size of the ring and the present-day SFR would constrain the epoch of collision, as discussed in the previous sections.

\section{Summary and conclusion}

We present hydrodynamical models of Cartwheel-like ring galaxies to explore the formation and evolution of the main structures, and the impact of large-scale dynamics on the enhanced star formation activity. Our main conclusions are as follows.
\begin{itemize}
\item The overall evolution of the system is independent of the initial intrinsic stability of the main galaxy's disc, at least in the range of $\qt$ we probed. Details in the star formation activity are however affected by the pre-collision existence and stability of gas structures and clouds, up to significant differences in the global SFR, even $100 \Myr$ after the encounter. It is likely that a disc yielding stronger structures (spirals, bars) before the collision than our models would respond differently to the encounter \citep[see][]{Athanassoula1997, Berentzen2003}.
\item The ring structure ploughs through the disc material as it radially expands at the roughly constant speed of $\approx 120 \kms$. This accretion of gas and stars is counterbalanced by mass-loss in the spokes, which in turn fuels the nuclear region. The ring is thus a short-lived feature that reaches it maximum mass $\approx 90 \Myr$ after the collision (when the ring's radius is $\approx 17 \kpc$), and becomes difficult to identify $\approx 110 \Myr$ later.
\item The passage of the companion galaxy also triggers the formation of a tidal vertical structure in the form of a diffuse annular, cylindrical shape above the ring. We predict a possible detection of such structures in the velocity field of real collisional ring galaxies, although the vertical velocity structure of this object encompasses several components that overlap with those of the ring itself and might be difficult to disentangle. The cylindrical shape could be easier to detect in the distribution of neutral hydrogen. Although no star formation is detected there in our models, because of too low gas densities, we speculate that other progenitor galaxies and/or other orbital setups could lead to enough gas accumulation to host a star formation activity, possibly in the form of tidal dwarf galaxy(ies). Such situation might well be what is observed, in particular in \ha, in AM~0644-741 (i.e. the Lindsay-Shapley galaxy).
\item A few Myr before the collision, the approaching companion galaxy tidally disrupts and destroys (at least some) gas overdensities, thus (at least partially) quenching the star formation activity across the target's disc. This activity resumes rapidly after the collision, in an enhanced manner. However, in the Kennicutt-Schmidt diagram, the system barely reaches the starburst regime.
\item The ring is the most actively star forming region during its first $100 \Myr$. After this phase, the flows of gas through the spokes transfer the star formation activity to the nuclear region. Short bursts in this activity occur when marginally stable gas clumps, that formed in the ring and survived along the spokes, fall into the nucleus. We find that spokes (and portions of spokes) with almost radial shapes yield a weaker shear than those with a smaller pitch angle, the former favouring the survival of gas clumps. Such sequential evolution of the location of the star formation activity could thus be used to estimate the dynamical stage of observed ring systems, and even date the epoch of encounter. That would likely provide strong constraints on the star formation histories of these galaxies.
\item The portion of the ring the furthest from the nucleus yields significantly more star formation than the other side, exactly as observed in the Cartwheel and e.g. AM~0644-741. This is linked to the weaker, and even compressive tidal field found there, while the other side hosts stronger, more destructive tides. The off-centred nucleus and dark matter halo thus have a stabilizing effect on the ISM of the ring.
\item After a destructive effect leading to star formation quenching when the companion approaches, the tidal field of the target galaxy becomes strongly compressive over kpc-scale volumes. The compressive nature of the tidal field is maintained in both the inner and outer rings after the companion has moved away from the target (with the notable exception of an azimuthal fraction at the latest stages).
\item This compression is transmitted to the turbulence of the ISM. Its compressive mode almost overcomes the otherwise dominant solenoidal (stirring) mode for a few $\sim 10 \Myr$ after the collision, but the turbulence rapidly goes back to energy equipartition of the two modes (i.e. solenoidal dominated). Such evolution translates into an excess of dense gas visible in the gas density PDF of the system, further leading to the mild but significant boost of the SFR shortly after the collision. This situation is similar to what is found in other galaxy interactions and mergers. The longer-term evolution of the star formation activity ($\gtrsim 50\mh 70\Myr$ after the collision) then rather proceeds like in normal star-forming discs, i.e. with a longer gas depletion time and a lower star formation efficiency, yet at a rate higher than that before the collision.
\item The interaction triggers an increase in the number of (massive) star clusters formed. While the stellar mass fraction found in clusters reaches a plateau (or even decreases) shortly after the collision due to tidal disruption, it jumps a few $\sim 10 \Myr$ later before slowing down or even stalling when the ring starts losing mass. The clusters formed are significantly more massive than those formed before the collision. The induced formation of the most massive clusters occurs $\approx 20 \Myr$ after the overall number of clusters rises. The most massive objects form in the ring where and when the tidal field and the turbulence are compressive and the shear is weak, during the first $\approx 100 \Myr$ after the collision. They then travel along the spokes and reach the nuclear region. Keeping in mind the limitations of our models, we speculate that the initial cluster mass function in the Cartwheel would be shallower than that of classical mergers like the Antennae, possibly because of the weaker effect of compressive turbulence. These differences call for a deeper and more systematic analysis of triggered star formation in interacting systems.
\end{itemize}

\section*{Acknowledgements}
We warmly thank Jean-Charles Lambert for his help with the visualisation of the simulation results and for making the tool glnemo2 public (\url{https://projets.lam.fr/projects/glnemo2}), Ramon Rey-Raposo and Fran\c coise Combes for discussions that helped improve the quality of this work, and the reviewer for their insightful comments. FR acknowledges support from the European Research Council through grant ERC-StG-335936. This work was granted access to the PRACE Research Infrastructure (under allocation 2540) and GENCI (under allocations 2192 and s030) resource \emph{Curie} hosted at the TGCC (France). 




\bibliographystyle{mnras}
\bibliography{biblio}

\begin{thebibliography}{}
\makeatletter
\relax
\def\mn@urlcharsother{\let\do\@makeother \do\$\do\&\do\#\do\^\do\_\do\%\do\~}
\def\mn@doi{\begingroup\mn@urlcharsother \@ifnextchar [ {\mn@doi@}
  {\mn@doi@[]}}
\def\mn@doi@[#1]#2{\def\@tempa{#1}\ifx\@tempa\@empty \href
  {http://dx.doi.org/#2} {doi:#2}\else \href {http://dx.doi.org/#2} {#1}\fi
  \endgroup}
\def\mn@eprint#1#2{\mn@eprint@#1:#2::\@nil}
\def\mn@eprint@arXiv#1{\href {http://arxiv.org/abs/#1} {{\tt arXiv:#1}}}
\def\mn@eprint@dblp#1{\href {http://dblp.uni-trier.de/rec/bibtex/#1.xml}
  {dblp:#1}}
\def\mn@eprint@#1:#2:#3:#4\@nil{\def\@tempa {#1}\def\@tempb {#2}\def\@tempc
  {#3}\ifx \@tempc \@empty \let \@tempc \@tempb \let \@tempb \@tempa \fi \ifx
  \@tempb \@empty \def\@tempb {arXiv}\fi \@ifundefined
  {mn@eprint@\@tempb}{\@tempb:\@tempc}{\expandafter \expandafter \csname
  mn@eprint@\@tempb\endcsname \expandafter{\@tempc}}}

\bibitem[\protect\citeauthoryear{{Agertz} \& {Kravtsov}}{{Agertz} \&
  {Kravtsov}}{2015}]{Agertz2015}
{Agertz} O.,  {Kravtsov} A.~V.,  2015, \mn@doi [\apj]
  {10.1088/0004-637X/804/1/18}, \href
  {http://adsabs.harvard.edu/abs/2015ApJ...804...18A} {804, 18}

\bibitem[\protect\citeauthoryear{{Agertz} \& {Kravtsov}}{{Agertz} \&
  {Kravtsov}}{2016}]{Agertz2016}
{Agertz} O.,  {Kravtsov} A.~V.,  2016, \mn@doi [\apj]
  {10.3847/0004-637X/824/2/79}, \href
  {http://adsabs.harvard.edu/abs/2016ApJ...824...79A} {824, 79}

\bibitem[\protect\citeauthoryear{{Amram}, {Mendes de Oliveira}, {Boulesteix}
  \& {Balkowski}}{{Amram} et~al.}{1998}]{Amram1998}
{Amram} P.,  {Mendes de Oliveira} C.,  {Boulesteix} J.,   {Balkowski} C.,
  1998, \aap, \href {http://adsabs.harvard.edu/abs/1998A%26A...330..881A} {330,
  881}

\bibitem[\protect\citeauthoryear{{Antonini}}{{Antonini}}{2013}]{Antonini2013}
{Antonini} F.,  2013, \mn@doi [\apj] {10.1088/0004-637X/763/1/62}, \href
  {http://adsabs.harvard.edu/abs/2013ApJ...763...62A} {763, 62}

\bibitem[\protect\citeauthoryear{{Appleton} \& {James}}{{Appleton} \&
  {James}}{1990}]{Appleton1990}
{Appleton} P.~N.,  {James} R.~A.,  1990, {Self-consistent simulations of ring
  galaxies.}.
pp 200--204

\bibitem[\protect\citeauthoryear{{Appleton} \& {Marston}}{{Appleton} \&
  {Marston}}{1997}]{Appleton1997}
{Appleton} P.~N.,  {Marston} A.~P.,  1997, \mn@doi [\aj] {10.1086/118245},
  \href {http://adsabs.harvard.edu/abs/1997AJ....113..201A} {113, 201}

\bibitem[\protect\citeauthoryear{{Appleton} \& {Struck-Marcell}}{{Appleton} \&
  {Struck-Marcell}}{1996}]{Appleton1996}
{Appleton} P.~N.,  {Struck-Marcell} C.,  1996, \fcp, \href
  {http://adsabs.harvard.edu/abs/1996FCPh...16..111A} {16, 111}

\bibitem[\protect\citeauthoryear{{Athanassoula} \& {Bosma}}{{Athanassoula} \&
  {Bosma}}{1985}]{Athanassoula1985}
{Athanassoula} E.,  {Bosma} A.,  1985, \mn@doi [\araa]
  {10.1146/annurev.aa.23.090185.001051}, \href
  {http://adsabs.harvard.edu/abs/1985ARA%26A..23..147A} {23, 147}

\bibitem[\protect\citeauthoryear{{Athanassoula}, {Puerari}  \&
  {Bosma}}{{Athanassoula} et~al.}{1997}]{Athanassoula1997}
{Athanassoula} E.,  {Puerari} I.,   {Bosma} A.,  1997, \mn@doi [\mnras]
  {10.1093/mnras/286.2.284}, \href
  {http://adsabs.harvard.edu/abs/1997MNRAS.286..284A} {286, 284}

\bibitem[\protect\citeauthoryear{{Barnes}}{{Barnes}}{2004}]{Barnes2004}
{Barnes} J.~E.,  2004, \mn@doi [\mnras] {10.1111/j.1365-2966.2004.07725.x},
  \href {http://adsabs.harvard.edu/abs/2004MNRAS.350..798B} {350, 798}

\bibitem[\protect\citeauthoryear{{Bastian}, {Trancho}, {Konstantopoulos}  \&
  {Miller}}{{Bastian} et~al.}{2009}]{Bastian2009}
{Bastian} N.,  {Trancho} G.,  {Konstantopoulos} I.~S.,   {Miller} B.~W.,  2009,
  \mn@doi [\apj] {10.1088/0004-637X/701/1/607}, \href
  {http://adsabs.harvard.edu/abs/2009ApJ...701..607B} {701, 607}

\bibitem[\protect\citeauthoryear{{Berentzen}, {Athanassoula}, {Heller}  \&
  {Fricke}}{{Berentzen} et~al.}{2003}]{Berentzen2003}
{Berentzen} I.,  {Athanassoula} E.,  {Heller} C.~H.,   {Fricke} K.~J.,  2003,
  \mn@doi [\mnras] {10.1046/j.1365-8711.2003.06417.x}, \href
  {http://adsabs.harvard.edu/abs/2003MNRAS.341..343B} {341, 343}

\bibitem[\protect\citeauthoryear{{Bosma}}{{Bosma}}{2000}]{Bosma2000}
{Bosma} A.,  2000, in {Valtonen} M.~J.,  {Flynn} C.,  eds,  Astronomical
  Society of the Pacific Conference Series Vol. 209, IAU Colloq. 174: Small
  Galaxy Groups. p.~255

\bibitem[\protect\citeauthoryear{{Bournaud} et~al.,}{{Bournaud}
  et~al.}{2007}]{Bournaud2007}
{Bournaud} F.,  et~al., 2007, \mn@doi [Science] {10.1126/science.1142114},
  \href {http://adsabs.harvard.edu/abs/2007Sci...316.1166B} {316, 1166}

\bibitem[\protect\citeauthoryear{{Bournaud}, {Elmegreen}, {Teyssier}, {Block}
  \& {Puerari}}{{Bournaud} et~al.}{2010}]{Bournaud2010b}
{Bournaud} F.,  {Elmegreen} B.~G.,  {Teyssier} R.,  {Block} D.~L.,   {Puerari}
  I.,  2010, \mn@doi [\mnras] {10.1111/j.1365-2966.2010.17370.x}, \href
  {http://adsabs.harvard.edu/abs/2010MNRAS.409.1088B} {409, 1088}

\bibitem[\protect\citeauthoryear{{Chabrier}}{{Chabrier}}{2003}]{Chabrier2003}
{Chabrier} G.,  2003, \mn@doi [\pasp] {10.1086/376392}, \href
  {http://adsabs.harvard.edu/abs/2003PASP..115..763C} {115, 763}

\bibitem[\protect\citeauthoryear{{Charmandaris}, {Laurent}, {Mirabel},
  {Gallais}, {Sauvage}, {Vigroux}, {Cesarsky}  \& {Appleton}}{{Charmandaris}
  et~al.}{1999}]{Charmandaris1999}
{Charmandaris} V.,  {Laurent} O.,  {Mirabel} I.~F.,  {Gallais} P.,  {Sauvage}
  M.,  {Vigroux} L.,  {Cesarsky} C.,   {Appleton} P.~N.,  1999, \aap, \href
  {http://adsabs.harvard.edu/abs/1999A%26A...341...69C} {341, 69}

\bibitem[\protect\citeauthoryear{{D'Onghia}, {Mapelli}  \& {Moore}}{{D'Onghia}
  et~al.}{2008}]{DOnghia2008}
{D'Onghia} E.,  {Mapelli} M.,   {Moore} B.,  2008, \mn@doi [\mnras]
  {10.1111/j.1365-2966.2008.13625.x}, \href
  {http://adsabs.harvard.edu/abs/2008MNRAS.389.1275D} {389, 1275}

\bibitem[\protect\citeauthoryear{{Daddi} et~al.,}{{Daddi}
  et~al.}{2010}]{Daddi2010}
{Daddi} E.,  et~al., 2010, \mn@doi [\apj] {10.1088/0004-637X/713/1/686}, \href
  {http://adsabs.harvard.edu/abs/2010ApJ...713..686D} {713, 686}

\bibitem[\protect\citeauthoryear{{Diemand}, {Madau}  \& {Moore}}{{Diemand}
  et~al.}{2005}]{Diemand2005}
{Diemand} J.,  {Madau} P.,   {Moore} B.,  2005, \mn@doi [\mnras]
  {10.1111/j.1365-2966.2005.09604.x}, \href
  {http://esoads.eso.org/abs/2005MNRAS.364..367D} {364, 367}

\bibitem[\protect\citeauthoryear{{Duc} \& {Renaud}}{{Duc} \&
  {Renaud}}{2013}]{Duc2013}
{Duc} P.-A.,  {Renaud} F.,  2013, in {Souchay} J.,  {Mathis} S.,   {Tokieda}
  T.,  eds,  Vol. 861, Lecture Notes in Physics, Berlin Springer Verlag. p.~327
  (\mn@eprint {arXiv} {1112.1922}), \mn@doi{10.1007/978-3-642-32961-6_9}

\bibitem[\protect\citeauthoryear{{Eisenstein} \& {Hut}}{{Eisenstein} \&
  {Hut}}{1998}]{Eisenstein1998}
{Eisenstein} D.~J.,  {Hut} P.,  1998, \mn@doi [\apj] {10.1086/305535}, \href
  {http://adsabs.harvard.edu/abs/1998ApJ...498..137E} {498, 137}

\bibitem[\protect\citeauthoryear{{Elmegreen} \& {Efremov}}{{Elmegreen} \&
  {Efremov}}{1997}]{ELmegreen1997}
{Elmegreen} B.~G.,  {Efremov} Y.~N.,  1997, \mn@doi [\apj] {10.1086/303966},
  \href
  {http://adsabs.harvard.edu/cgi-bin/nph-bib_query?bibcode=1997ApJ...480..235E&db_key=AST}
  {480, 235}

\bibitem[\protect\citeauthoryear{{Elmegreen}, {Kaufman}, {Brinks}, {Elmegreen}
  \& {Sundin}}{{Elmegreen} et~al.}{1995}]{Elmegreen1995b}
{Elmegreen} D.~M.,  {Kaufman} M.,  {Brinks} E.,  {Elmegreen} B.~G.,   {Sundin}
  M.,  1995, \mn@doi [\apj] {10.1086/176374}, \href
  {http://adsabs.harvard.edu/abs/1995ApJ...453..100E} {453, 100}

\bibitem[\protect\citeauthoryear{{Federrath}, {Klessen}  \&
  {Schmidt}}{{Federrath} et~al.}{2008}]{Federrath2008}
{Federrath} C.,  {Klessen} R.~S.,   {Schmidt} W.,  2008, \mn@doi [\apjl]
  {10.1086/595280}, \href {http://adsabs.harvard.edu/abs/2008ApJ...688L..79F}
  {688, L79}

\bibitem[\protect\citeauthoryear{{Fensch} et~al.,}{{Fensch}
  et~al.}{2017}]{Fensch2017}
{Fensch} J.,  et~al., 2017, \mn@doi [\mnras] {10.1093/mnras/stw2920}, \href
  {http://adsabs.harvard.edu/abs/2017MNRAS.465.1934F} {465, 1934}

\bibitem[\protect\citeauthoryear{{Fiacconi}, {Mapelli}, {Ripamonti}  \&
  {Colpi}}{{Fiacconi} et~al.}{2012}]{Fiacconi2012}
{Fiacconi} D.,  {Mapelli} M.,  {Ripamonti} E.,   {Colpi} M.,  2012, \mn@doi
  [\mnras] {10.1111/j.1365-2966.2012.21566.x}, \href
  {http://adsabs.harvard.edu/abs/2012MNRAS.425.2255F} {425, 2255}

\bibitem[\protect\citeauthoryear{{Fosbury} \& {Hawarden}}{{Fosbury} \&
  {Hawarden}}{1977}]{Fosbury1977}
{Fosbury} R.~A.~E.,  {Hawarden} T.~G.,  1977, \mn@doi [\mnras]
  {10.1093/mnras/178.3.473}, \href
  {http://adsabs.harvard.edu/abs/1977MNRAS.178..473F} {178, 473}

\bibitem[\protect\citeauthoryear{{Gao}, {Wang}, {Appleton}  \& {Lucas}}{{Gao}
  et~al.}{2003}]{Gao2003}
{Gao} Y.,  {Wang} Q.~D.,  {Appleton} P.~N.,   {Lucas} R.~A.,  2003, \mn@doi
  [\apjl] {10.1086/379598}, \href
  {http://adsabs.harvard.edu/abs/2003ApJ...596L.171G} {596, L171}

\bibitem[\protect\citeauthoryear{{Genzel}, {Tacconi}, {Gracia-Carpio},
  {Sternberg}, {Cooper}, {Shapiro}, {Bolatto}  \& {et al.}}{{Genzel}
  et~al.}{2010}]{Genzel2010}
{Genzel} R.,  {Tacconi} L.~J.,  {Gracia-Carpio} J.,  {Sternberg} A.,  {Cooper}
  M.~C.,  {Shapiro} K.,  {Bolatto} A.,   {et al.} 2010, \mn@doi [\mnras]
  {10.1111/j.1365-2966.2010.16969.x}, \href
  {http://adsabs.harvard.edu/abs/2010MNRAS.407.2091G} {407, 2091}

\bibitem[\protect\citeauthoryear{{Georgakakis}, {Forbes}  \&
  {Norris}}{{Georgakakis} et~al.}{2000}]{Georgakakis2000}
{Georgakakis} A.,  {Forbes} D.~A.,   {Norris} R.~P.,  2000, \mn@doi [\mnras]
  {10.1046/j.1365-8711.2000.03709.x}, \href
  {http://adsabs.harvard.edu/abs/2000MNRAS.318..124G} {318, 124}

\bibitem[\protect\citeauthoryear{{Gerber} \& {Lamb}}{{Gerber} \&
  {Lamb}}{1994}]{Gerber1994}
{Gerber} R.~A.,  {Lamb} S.~A.,  1994, \mn@doi [\apj] {10.1086/174511}, \href
  {http://adsabs.harvard.edu/abs/1994ApJ...431..604G} {431, 604}

\bibitem[\protect\citeauthoryear{{Gerber}, {Lamb}  \& {Balsara}}{{Gerber}
  et~al.}{1992}]{Gerber1992}
{Gerber} R.~A.,  {Lamb} S.~A.,   {Balsara} D.~S.,  1992, \mn@doi [\apjl]
  {10.1086/186604}, \href {http://adsabs.harvard.edu/abs/1992ApJ...399L..51G}
  {399, L51}

\bibitem[\protect\citeauthoryear{{Gerber}, {Lamb}  \& {Balsara}}{{Gerber}
  et~al.}{1996}]{Gerber1996}
{Gerber} R.~A.,  {Lamb} S.~A.,   {Balsara} D.~S.,  1996, \mn@doi [\mnras]
  {10.1093/mnras/278.2.345}, \href
  {http://adsabs.harvard.edu/abs/1996MNRAS.278..345G} {278, 345}

\bibitem[\protect\citeauthoryear{{Ghosh} \& {Mapelli}}{{Ghosh} \&
  {Mapelli}}{2008}]{Ghosh2008}
{Ghosh} K.~K.,  {Mapelli} M.,  2008, \mn@doi [\mnras]
  {10.1111/j.1745-3933.2008.00456.x}, \href
  {http://adsabs.harvard.edu/abs/2008MNRAS.386L..38G} {386, L38}

\bibitem[\protect\citeauthoryear{{Gieles} \& {Renaud}}{{Gieles} \&
  {Renaud}}{2016}]{Gieles2016}
{Gieles} M.,  {Renaud} F.,  2016, \mn@doi [\mnras] {10.1093/mnrasl/slw163},
  \href {http://adsabs.harvard.edu/abs/2016MNRAS.463L.103G} {463, L103}

\bibitem[\protect\citeauthoryear{{Gieles}, {Athanassoula}  \& {Portegies
  Zwart}}{{Gieles} et~al.}{2007}]{Gieles2007}
{Gieles} M.,  {Athanassoula} E.,   {Portegies Zwart} S.~F.,  2007, \mn@doi
  [\mnras] {10.1111/j.1365-2966.2007.11477.x}, \href
  {http://adsabs.harvard.edu/abs/2007MNRAS.376..809G} {376, 809}

\bibitem[\protect\citeauthoryear{{Graham}}{{Graham}}{1974}]{Graham1974}
{Graham} J.~A.,  1974, The Observatory, \href
  {http://adsabs.harvard.edu/abs/1974Obs....94..290G} {94, 290}

\bibitem[\protect\citeauthoryear{{Grisdale}, {Agertz}, {Romeo}, {Renaud}  \&
  {Read}}{{Grisdale} et~al.}{2017}]{Grisdale2017}
{Grisdale} K.,  {Agertz} O.,  {Romeo} A.~B.,  {Renaud} F.,   {Read} J.~I.,
  2017, \mn@doi [\mnras] {10.1093/mnras/stw3133}, \href
  {http://adsabs.harvard.edu/abs/2017MNRAS.466.1093G} {466, 1093}

\bibitem[\protect\citeauthoryear{{Guillard}, {Emsellem}  \&
  {Renaud}}{{Guillard} et~al.}{2016}]{Guillard2016}
{Guillard} N.,  {Emsellem} E.,   {Renaud} F.,  2016, \mn@doi [\mnras]
  {10.1093/mnras/stw1570}, \href
  {http://adsabs.harvard.edu/abs/2016MNRAS.461.3620G} {461, 3620}

\bibitem[\protect\citeauthoryear{{Heggie} \& {Hut}}{{Heggie} \&
  {Hut}}{2003}]{Heggie2003}
{Heggie} D.,  {Hut} P.,  2003, {The Gravitational Million-Body Problem: A
  Multidisciplinary Approach to Star Cluster Dynamics, by Douglas Heggie and
  Piet Hut}.
Cambridge University Press, 2003, 372 pp.

\bibitem[\protect\citeauthoryear{{Hennebelle} \& {Falgarone}}{{Hennebelle} \&
  {Falgarone}}{2012}]{Hennebelle2012}
{Hennebelle} P.,  {Falgarone} E.,  2012, \mn@doi [\aapr]
  {10.1007/s00159-012-0055-y}, \href
  {http://adsabs.harvard.edu/abs/2012A%26ARv..20...55H} {20, 55}

\bibitem[\protect\citeauthoryear{{Hernquist} \& {Weil}}{{Hernquist} \&
  {Weil}}{1993}]{Hernquist1993}
{Hernquist} L.,  {Weil} M.~L.,  1993, \mnras, \href
  {http://adsabs.harvard.edu/abs/1993MNRAS.261..804H} {261, 804}

\bibitem[\protect\citeauthoryear{{Higdon}}{{Higdon}}{1995}]{Higdon1995}
{Higdon} J.~L.,  1995, \mn@doi [\apj] {10.1086/176602}, \href
  {http://adsabs.harvard.edu/abs/1995ApJ...455..524H} {455, 524}

\bibitem[\protect\citeauthoryear{{Higdon}}{{Higdon}}{1996}]{Higdon1996}
{Higdon} J.~L.,  1996, \mn@doi [\apj] {10.1086/177599}, \href
  {http://adsabs.harvard.edu/abs/1996ApJ...467..241H} {467, 241}

\bibitem[\protect\citeauthoryear{{Higdon}, {Higdon}  \& {Rand}}{{Higdon}
  et~al.}{2011}]{Higdon2011}
{Higdon} J.~L.,  {Higdon} S.~J.~U.,   {Rand} R.~J.,  2011, \mn@doi [\apj]
  {10.1088/0004-637X/739/2/97}, \href
  {http://adsabs.harvard.edu/abs/2011ApJ...739...97H} {739, 97}

\bibitem[\protect\citeauthoryear{{Higdon}, {Higdon}, {Mart{\'{\i}}n Ruiz}  \&
  {Rand}}{{Higdon} et~al.}{2015}]{Higdon2015}
{Higdon} J.~L.,  {Higdon} S.~J.~U.,  {Mart{\'{\i}}n Ruiz} S.,   {Rand} R.~J.,
  2015, \mn@doi [\apjl] {10.1088/2041-8205/814/1/L1}, \href
  {http://adsabs.harvard.edu/abs/2015ApJ...814L...1H} {814, L1}

\bibitem[\protect\citeauthoryear{{Horellou} \& {Combes}}{{Horellou} \&
  {Combes}}{2001}]{Horellou2001}
{Horellou} C.,  {Combes} F.,  2001, \mn@doi [\apss] {10.1023/A:1017524632342},
  \href {http://adsabs.harvard.edu/abs/2001Ap%26SS.276.1141H} {276, 1141}

\bibitem[\protect\citeauthoryear{{Horellou}, {Charmandaris}, {Combes},
  {Appleton}, {Casoli}  \& {Mirabel}}{{Horellou} et~al.}{1998}]{Horellou1998}
{Horellou} C.,  {Charmandaris} V.,  {Combes} F.,  {Appleton} P.~N.,  {Casoli}
  F.,   {Mirabel} I.~F.,  1998, \aap, \href
  {http://adsabs.harvard.edu/abs/1998A%26A...340L..51H} {340, L51}

\bibitem[\protect\citeauthoryear{{Huang} \& {Stewart}}{{Huang} \&
  {Stewart}}{1988}]{Huang1988}
{Huang} S.-N.,  {Stewart} P.,  1988, \aap, \href
  {http://adsabs.harvard.edu/abs/1988A%26A...197...14H} {197, 14}

\bibitem[\protect\citeauthoryear{{Iovino}}{{Iovino}}{2002}]{Iovino2002}
{Iovino} A.,  2002, \mn@doi [\aj] {10.1086/343059}, \href
  {http://adsabs.harvard.edu/abs/2002AJ....124.2471I} {124, 2471}

\bibitem[\protect\citeauthoryear{{Irwin}}{{Irwin}}{1994}]{Irwin1994}
{Irwin} J.~A.,  1994, \mn@doi [\apj] {10.1086/174349}, \href
  {http://adsabs.harvard.edu/abs/1994ApJ...429..618I} {429, 618}

\bibitem[\protect\citeauthoryear{{Jog} \& {Solomon}}{{Jog} \&
  {Solomon}}{1992}]{Jog1992}
{Jog} C.~J.,  {Solomon} P.~M.,  1992, \mn@doi [\apj] {10.1086/171067}, \href
  {http://adsabs.harvard.edu/abs/1992ApJ...387..152J} {387, 152}

\bibitem[\protect\citeauthoryear{{Kennicutt}}{{Kennicutt}}{1998}]{Kennicutt1998b}
{Kennicutt} R.~C.,  1998, \mn@doi [\apj] {10.1086/305588}, \href
  {http://adsabs.harvard.edu/cgi-bin/nph-bib_query?bibcode=1998ApJ...498..541K&db_key=AST}
  {498, 541}

\bibitem[\protect\citeauthoryear{{Kim} \& {Ostriker}}{{Kim} \&
  {Ostriker}}{2015}]{Kim2015}
{Kim} C.-G.,  {Ostriker} E.~C.,  2015, \mn@doi [\apj]
  {10.1088/0004-637X/802/2/99}, \href
  {http://adsabs.harvard.edu/abs/2015ApJ...802...99K} {802, 99}

\bibitem[\protect\citeauthoryear{{Klessen} \& {Hennebelle}}{{Klessen} \&
  {Hennebelle}}{2010}]{Klessen2010}
{Klessen} R.~S.,  {Hennebelle} P.,  2010, \mn@doi [\aap]
  {10.1051/0004-6361/200913780}, \href
  {http://adsabs.harvard.edu/abs/2010A%26A...520A..17K} {520, A17}

\bibitem[\protect\citeauthoryear{{Kravtsov} \& {Gnedin}}{{Kravtsov} \&
  {Gnedin}}{2005}]{Kravtsov2005}
{Kravtsov} A.~V.,  {Gnedin} O.~Y.,  2005, \mn@doi [\apj] {10.1086/428636},
  \href {http://adsabs.harvard.edu/abs/2005ApJ...623..650K} {623, 650}

\bibitem[\protect\citeauthoryear{{Krumholz} \& {Burkhart}}{{Krumholz} \&
  {Burkhart}}{2016}]{Krumholz2016}
{Krumholz} M.~R.,  {Burkhart} B.,  2016, \mn@doi [\mnras]
  {10.1093/mnras/stw434}, \href
  {http://adsabs.harvard.edu/abs/2016MNRAS.458.1671K} {458, 1671}

\bibitem[\protect\citeauthoryear{{Lardo}, {Cabrera-Ziri}, {Davies}  \&
  {Bastian}}{{Lardo} et~al.}{2017}]{Lardo2017}
{Lardo} C.,  {Cabrera-Ziri} I.,  {Davies} B.,   {Bastian} N.,  2017, \mn@doi
  [\mnras] {10.1093/mnras/stx628}, \href
  {http://adsabs.harvard.edu/abs/2017MNRAS.468.2482L} {468, 2482}

\bibitem[\protect\citeauthoryear{{Lavery}, {Remijan}, {Charmandaris}, {Hayes}
  \& {Ring}}{{Lavery} et~al.}{2004}]{Lavery2004}
{Lavery} R.~J.,  {Remijan} A.,  {Charmandaris} V.,  {Hayes} R.~D.,   {Ring}
  A.~A.,  2004, \mn@doi [\apj] {10.1086/422420}, \href
  {http://adsabs.harvard.edu/abs/2004ApJ...612..679L} {612, 679}

\bibitem[\protect\citeauthoryear{{Lynds} \& {Toomre}}{{Lynds} \&
  {Toomre}}{1976}]{Lynds1976}
{Lynds} R.,  {Toomre} A.,  1976, \mn@doi [\apj] {10.1086/154730}, \href
  {http://adsabs.harvard.edu/abs/1976ApJ...209..382L} {209, 382}

\bibitem[\protect\citeauthoryear{{Mac Low} \& {Klessen}}{{Mac Low} \&
  {Klessen}}{2004}]{MacLow2004}
{Mac Low} M.-M.,  {Klessen} R.~S.,  2004, \mn@doi [Reviews of Modern Physics]
  {10.1103/RevModPhys.76.125}, \href
  {http://adsabs.harvard.edu/abs/2004RvMP...76..125M} {76, 125}

\bibitem[\protect\citeauthoryear{{Madore}, {Nelson}  \& {Petrillo}}{{Madore}
  et~al.}{2009}]{Madore2009}
{Madore} B.~F.,  {Nelson} E.,   {Petrillo} K.,  2009, \mn@doi [\apjs]
  {10.1088/0067-0049/181/2/572}, \href
  {http://adsabs.harvard.edu/abs/2009ApJS..181..572M} {181, 572}

\bibitem[\protect\citeauthoryear{{Mapelli} \& {Mayer}}{{Mapelli} \&
  {Mayer}}{2012}]{Mapelli2012}
{Mapelli} M.,  {Mayer} L.,  2012, \mn@doi [\mnras]
  {10.1111/j.1365-2966.2011.20098.x}, \href
  {http://adsabs.harvard.edu/abs/2012MNRAS.420.1158M} {420, 1158}

\bibitem[\protect\citeauthoryear{{Mapelli}, {Moore}, {Ripamonti}, {Mayer},
  {Colpi}  \& {Giordano}}{{Mapelli} et~al.}{2008}]{Mapelli2008}
{Mapelli} M.,  {Moore} B.,  {Ripamonti} E.,  {Mayer} L.,  {Colpi} M.,
  {Giordano} L.,  2008, \mn@doi [\mnras] {10.1111/j.1365-2966.2007.12650.x},
  \href {http://adsabs.harvard.edu/abs/2008MNRAS.383.1223M} {383, 1223}

\bibitem[\protect\citeauthoryear{{Marcum}, {Appleton}  \& {Higdon}}{{Marcum}
  et~al.}{1992}]{Marcum1992}
{Marcum} P.~M.,  {Appleton} P.~N.,   {Higdon} J.~L.,  1992, \mn@doi [\apj]
  {10.1086/171902}, \href {http://adsabs.harvard.edu/abs/1992ApJ...399...57M}
  {399, 57}

\bibitem[\protect\citeauthoryear{{Matzner}}{{Matzner}}{2002}]{Matzner2002}
{Matzner} C.~D.,  2002, \mn@doi [\apj] {10.1086/338030}, \href
  {http://adsabs.harvard.edu/abs/2002ApJ...566..302M} {566, 302}

\bibitem[\protect\citeauthoryear{{Mayya}, {Bizyaev}, {Romano}, {Garcia-Barreto}
   \& {Vorobyov}}{{Mayya} et~al.}{2005}]{Mayya2005}
{Mayya} Y.~D.,  {Bizyaev} D.,  {Romano} R.,  {Garcia-Barreto} J.~A.,
  {Vorobyov} E.~I.,  2005, \mn@doi [\apjl] {10.1086/428400}, \href
  {http://adsabs.harvard.edu/abs/2005ApJ...620L..35M} {620, L35}

\bibitem[\protect\citeauthoryear{{Meurer} \& {et al.}}{{Meurer} \& {et
  al.}}{2006}]{Meurer2006}
{Meurer} G.~R.,  {et al.} 2006, \mn@doi [\apjs] {10.1086/504685}, \href
  {http://adsabs.harvard.edu/abs/2006ApJS..165..307M} {165, 307}

\bibitem[\protect\citeauthoryear{{Michel-Dansac} et~al.,}{{Michel-Dansac}
  et~al.}{2010}]{Michel2010}
{Michel-Dansac} L.,  et~al., 2010, \mn@doi [\apjl]
  {10.1088/2041-8205/717/2/L143}, \href
  {http://adsabs.harvard.edu/abs/2010ApJ...717L.143M} {717, L143}

\bibitem[\protect\citeauthoryear{{Milosavljevi{\'c}}}{{Milosavljevi{\'c}}}{2004}]{Milosavljevic2004}
{Milosavljevi{\'c}} M.,  2004, \mn@doi [\apjl] {10.1086/420696}, \href
  {http://adsabs.harvard.edu/abs/2004ApJ...605L..13M} {605, L13}

\bibitem[\protect\citeauthoryear{{Ossenkopf} \& {Mac Low}}{{Ossenkopf} \& {Mac
  Low}}{2002}]{Ossenkopf2002}
{Ossenkopf} V.,  {Mac Low} M.-M.,  2002, \mn@doi [\aap]
  {10.1051/0004-6361:20020629}, \href
  {http://adsabs.harvard.edu/abs/2002A%26A...390..307O} {390, 307}

\bibitem[\protect\citeauthoryear{{Pardy}, {D'Onghia}, {Athanassoula}, {Wilcots}
   \& {Sheth}}{{Pardy} et~al.}{2016}]{Pardy2016}
{Pardy} S.~A.,  {D'Onghia} E.,  {Athanassoula} E.,  {Wilcots} E.~M.,   {Sheth}
  K.,  2016, \mn@doi [\apj] {10.3847/0004-637X/827/2/149}, \href
  {http://adsabs.harvard.edu/abs/2016ApJ...827..149P} {827, 149}

\bibitem[\protect\citeauthoryear{{Pellerin}, {Meurer}, {Bekki}, {Elmegreen},
  {Wong}  \& {Knezek}}{{Pellerin} et~al.}{2010}]{Pellerin2010}
{Pellerin} A.,  {Meurer} G.~R.,  {Bekki} K.,  {Elmegreen} D.~M.,  {Wong} O.~I.,
    {Knezek} P.~M.,  2010, \mn@doi [\aj] {10.1088/0004-6256/139/4/1369}, \href
  {http://adsabs.harvard.edu/abs/2010AJ....139.1369P} {139, 1369}

\bibitem[\protect\citeauthoryear{{Plummer}}{{Plummer}}{1911}]{Plummer1911}
{Plummer} H.~C.,  1911, \mnras, \href
  {http://adsabs.harvard.edu/cgi-bin/nph-bib_query?bibcode=1911MNRAS..71..460P&db_key=AST}
  {71, 460}

\bibitem[\protect\citeauthoryear{{Portegies Zwart}, {McMillan}  \&
  {Gieles}}{{Portegies Zwart} et~al.}{2010}]{Portegies2010}
{Portegies Zwart} S.~F.,  {McMillan} S.~L.~W.,   {Gieles} M.,  2010, \mn@doi
  [\araa] {10.1146/annurev-astro-081309-130834}, \href
  {http://adsabs.harvard.edu/abs/2010ARA%26A..48..431P} {48, 431}

\bibitem[\protect\citeauthoryear{{Prestwich} et~al.,}{{Prestwich}
  et~al.}{2012}]{Prestwich2012}
{Prestwich} A.~H.,  et~al., 2012, \mn@doi [\apj] {10.1088/0004-637X/747/2/150},
  \href {http://adsabs.harvard.edu/abs/2012ApJ...747..150P} {747, 150}

\bibitem[\protect\citeauthoryear{{Renaud}, {Boily}, {Naab}  \&
  {Theis}}{{Renaud} et~al.}{2009}]{Renaud2009}
{Renaud} F.,  {Boily} C.~M.,  {Naab} T.,   {Theis} C.,  2009, \mn@doi [\apj]
  {10.1088/0004-637X/706/1/67}, \href
  {http://adsabs.harvard.edu/abs/2009ApJ...706...67R} {706, 67}

\bibitem[\protect\citeauthoryear{{Renaud}, {Kraljic}  \& {Bournaud}}{{Renaud}
  et~al.}{2012}]{Renaud2012}
{Renaud} F.,  {Kraljic} K.,   {Bournaud} F.,  2012, \mn@doi [\apjl]
  {10.1088/2041-8205/760/1/L16}, \href
  {http://adsabs.harvard.edu/abs/2012ApJ...760L..16R} {760, L16}

\bibitem[\protect\citeauthoryear{{Renaud} et~al.,}{{Renaud}
  et~al.}{2013}]{Renaud2013b}
{Renaud} F.,  et~al., 2013, \mn@doi [\mnras] {10.1093/mnras/stt1698}, \href
  {http://adsabs.harvard.edu/abs/2013MNRAS.436.1836R} {436, 1836}

\bibitem[\protect\citeauthoryear{{Renaud}, {Bournaud}, {Kraljic}  \&
  {Duc}}{{Renaud} et~al.}{2014a}]{Renaud2014b}
{Renaud} F.,  {Bournaud} F.,  {Kraljic} K.,   {Duc} P.-A.,  2014a, \mn@doi
  [\mnras] {10.1093/mnrasl/slu050}, \href
  {http://adsabs.harvard.edu/abs/2014MNRAS.442L..33R} {442, L33}

\bibitem[\protect\citeauthoryear{{Renaud}, {Bournaud}, {Emsellem}, {Elmegreen}
  \& {Teyssier}}{{Renaud} et~al.}{2014b}]{Renaud2014}
{Renaud} F.,  {Bournaud} F.,  {Emsellem} E.,  {Elmegreen} B.,   {Teyssier} R.,
  2014b, in {Seigar} M.~S.,  {Treuthardt} P.,  eds,  Astronomical Society of
  the Pacific Conference Series Vol. 480, Structure and Dynamics of Disk
  Galaxies. p.~247 (\mn@eprint {arXiv} {1310.0082})

\bibitem[\protect\citeauthoryear{{Renaud}, {Bournaud}  \& {Duc}}{{Renaud}
  et~al.}{2015}]{Renaud2015}
{Renaud} F.,  {Bournaud} F.,   {Duc} P.-A.,  2015, \mn@doi [\mnras]
  {10.1093/mnras/stu2208}, \href
  {http://adsabs.harvard.edu/abs/2015MNRAS.446.2038R} {446, 2038}

\bibitem[\protect\citeauthoryear{{Renaud}, {Famaey}  \& {Kroupa}}{{Renaud}
  et~al.}{2016}]{Renaud2016}
{Renaud} F.,  {Famaey} B.,   {Kroupa} P.,  2016, \mn@doi [\mnras]
  {10.1093/mnras/stw2331}, \href
  {http://adsabs.harvard.edu/abs/2016MNRAS.463.3637R} {463, 3637}

\bibitem[\protect\citeauthoryear{{Renaud}, {Agertz}  \& {Gieles}}{{Renaud}
  et~al.}{2017}]{Renaud2017}
{Renaud} F.,  {Agertz} O.,   {Gieles} M.,  2017, \mn@doi [\mnras]
  {10.1093/mnras/stw2969}, \href
  {http://adsabs.harvard.edu/abs/2017MNRAS.465.3622R} {465, 3622}

\bibitem[\protect\citeauthoryear{{Romano}, {Mayya}  \& {Vorobyov}}{{Romano}
  et~al.}{2008}]{Romano2008}
{Romano} R.,  {Mayya} Y.~D.,   {Vorobyov} E.~I.,  2008, \mn@doi [\aj]
  {10.1088/0004-6256/136/3/1259}, \href
  {http://adsabs.harvard.edu/abs/2008AJ....136.1259R} {136, 1259}

\bibitem[\protect\citeauthoryear{{Schreiber} et~al.,}{{Schreiber}
  et~al.}{2015}]{Schreiber2015}
{Schreiber} C.,  et~al., 2015, \mn@doi [\aap] {10.1051/0004-6361/201425017},
  \href {http://adsabs.harvard.edu/abs/2015A%26A...575A..74S} {575, A74}

\bibitem[\protect\citeauthoryear{{Scudder}, {Ellison}, {Torrey}, {Patton}  \&
  {Mendel}}{{Scudder} et~al.}{2012}]{Scudder2012}
{Scudder} J.~M.,  {Ellison} S.~L.,  {Torrey} P.,  {Patton} D.~R.,   {Mendel}
  J.~T.,  2012, \mn@doi [\mnras] {10.1111/j.1365-2966.2012.21749.x}, \href
  {http://adsabs.harvard.edu/abs/2012MNRAS.426..549S} {426, 549}

\bibitem[\protect\citeauthoryear{{Smith}, {Lane}, {Conn}  \&
  {Fellhauer}}{{Smith} et~al.}{2012}]{Smith2012}
{Smith} R.,  {Lane} R.~R.,  {Conn} B.~C.,   {Fellhauer} M.,  2012, \mn@doi
  [\mnras] {10.1111/j.1365-2966.2012.20911.x}, \href
  {http://adsabs.harvard.edu/abs/2012MNRAS.423..543S} {423, 543}

\bibitem[\protect\citeauthoryear{{Struck-Marcell} \&
  {Appleton}}{{Struck-Marcell} \& {Appleton}}{1987}]{Struck1987}
{Struck-Marcell} C.,  {Appleton} P.~N.,  1987, \mn@doi [\apj] {10.1086/165846},
  \href {http://adsabs.harvard.edu/abs/1987ApJ...323..480S} {323, 480}

\bibitem[\protect\citeauthoryear{{Struck-Marcell} \& {Higdon}}{{Struck-Marcell}
  \& {Higdon}}{1993}]{Struck1993}
{Struck-Marcell} C.,  {Higdon} J.~L.,  1993, \mn@doi [\apj] {10.1086/172811},
  \href {http://adsabs.harvard.edu/abs/1993ApJ...411..108S} {411, 108}

\bibitem[\protect\citeauthoryear{{Struck-Marcell} \& {Lotan}}{{Struck-Marcell}
  \& {Lotan}}{1990}]{Struck1990}
{Struck-Marcell} C.,  {Lotan} P.,  1990, \mn@doi [\apj] {10.1086/168965}, \href
  {http://adsabs.harvard.edu/abs/1990ApJ...358...99S} {358, 99}

\bibitem[\protect\citeauthoryear{{Struck}, {Appleton}, {Borne}  \&
  {Lucas}}{{Struck} et~al.}{1996}]{Struck1996}
{Struck} C.,  {Appleton} P.~N.,  {Borne} K.~D.,   {Lucas} R.~A.,  1996, \mn@doi
  [\aj] {10.1086/118148}, \href
  {http://adsabs.harvard.edu/abs/1996AJ....112.1868S} {112, 1868}

\bibitem[\protect\citeauthoryear{{Teyssier}}{{Teyssier}}{2002}]{Teyssier2002}
{Teyssier} R.,  2002, \mn@doi [\aap] {10.1051/0004-6361:20011817}, \href
  {http://adsabs.harvard.edu/abs/2002A%26A...385..337T} {385, 337}

\bibitem[\protect\citeauthoryear{{Theys} \& {Spiegel}}{{Theys} \&
  {Spiegel}}{1977}]{Theys1977}
{Theys} J.~C.,  {Spiegel} E.~A.,  1977, \mn@doi [\apj] {10.1086/155084}, \href
  {http://adsabs.harvard.edu/abs/1977ApJ...212..616T} {212, 616}

\bibitem[\protect\citeauthoryear{{Toomre}}{{Toomre}}{1964}]{Toomre1964}
{Toomre} A.,  1964, \mn@doi [\apj] {10.1086/147861}, \href
  {http://adsabs.harvard.edu/abs/1964ApJ...139.1217T} {139, 1217}

\bibitem[\protect\citeauthoryear{{Toomre}}{{Toomre}}{1978}]{Toomre1978}
{Toomre} A.,  1978, in {Longair} M.~S.,  {Einasto} J.,  eds,  IAU Symposium
  Vol. 79, Large Scale Structures in the Universe. pp 109--116

\bibitem[\protect\citeauthoryear{{Toomre} \& {Toomre}}{{Toomre} \&
  {Toomre}}{1972}]{Toomre1972}
{Toomre} A.,  {Toomre} J.,  1972, \apj, \href
  {http://adsabs.harvard.edu/cgi-bin/nph-bib_query?bibcode=1972ApJ...178..623T&db_key=AST}
  {178, 623}

\bibitem[\protect\citeauthoryear{{Tremaine}, {Ostriker}  \&
  {Spitzer}}{{Tremaine} et~al.}{1975}]{Tremaine1975}
{Tremaine} S.~D.,  {Ostriker} J.~P.,   {Spitzer} Jr. L.,  1975, \mn@doi [\apj]
  {10.1086/153422}, \href {http://adsabs.harvard.edu/abs/1975ApJ...196..407T}
  {196, 407}

\bibitem[\protect\citeauthoryear{{Vorobyov}}{{Vorobyov}}{2003}]{Vorobyov2003}
{Vorobyov} E.~I.,  2003, \mn@doi [\aap] {10.1051/0004-6361:20031019}, \href
  {http://adsabs.harvard.edu/abs/2003A%26A...407..913V} {407, 913}

\bibitem[\protect\citeauthoryear{{Whitmore} \& {Schweizer}}{{Whitmore} \&
  {Schweizer}}{1995}]{Whitmore1995}
{Whitmore} B.~C.,  {Schweizer} F.,  1995, \mn@doi [\aj] {10.1086/117334}, \href
  {http://adsabs.harvard.edu/cgi-bin/nph-bib_query?bibcode=1995AJ....109..960W&db_key=AST}
  {109, 960}

\bibitem[\protect\citeauthoryear{{Wolter} \& {Trinchieri}}{{Wolter} \&
  {Trinchieri}}{2004}]{Wolter2004}
{Wolter} A.,  {Trinchieri} G.,  2004, \mn@doi [\aap]
  {10.1051/0004-6361:20047110}, \href
  {http://adsabs.harvard.edu/abs/2004A%26A...426..787W} {426, 787}

\bibitem[\protect\citeauthoryear{{Wolter}, {Trinchieri}  \& {Iovino}}{{Wolter}
  et~al.}{1999}]{Wolter1999}
{Wolter} A.,  {Trinchieri} G.,   {Iovino} A.,  1999, \aap, \href
  {http://adsabs.harvard.edu/abs/1999A%26A...342...41W} {342, 41}

\bibitem[\protect\citeauthoryear{{Wong} \& {et al.}}{{Wong} \& {et
  al.}}{2006}]{Wong2006}
{Wong} O.~I.,  {et al.} 2006, \mn@doi [\mnras]
  {10.1111/j.1365-2966.2006.10589.x}, \href
  {http://adsabs.harvard.edu/abs/2006MNRAS.370.1607W} {370, 1607}

\bibitem[\protect\citeauthoryear{{Zwicky}}{{Zwicky}}{1941}]{Zwicky1941}
{Zwicky} F.,  1941, {Hydrodynamics and the Structure of Stellar Systems}.
p.~137

\makeatother
\end{thebibliography}

\end{document}